\documentclass[useAMS,usenatbib]{mn2e}
\usepackage[a4paper,margin=2cm]{geometry}
\usepackage[pdftex]{graphicx}
\pdfoutput=1
\usepackage{color}
\usepackage{aas_macros}
\usepackage{lastpage}
\usepackage{comment}
\usepackage{amsmath}
\usepackage{type1cm}
\usepackage{eso-pic}
\usepackage{url}

\def\lsim{~\rlap{$<$}{\lower 1.0ex\hbox{$\sim$}}}

\def\gsim{~\rlap{$>$}{\lower 1.0ex\hbox{$\sim$}}}

\def\lesssim{\mathrel{\hbox{\rlap{\hbox{\lower4pt\hbox{$\sim$}}}\hbox{$<$}}}}
\def\gtrsim{\mathrel{\hbox{\rlap{\hbox{\lower4pt\hbox{$\sim$}}}\hbox{$>$}}}}
\newcommand{\ltaraw}{$\; \buildrel < \over \sim \;$}
\newcommand{\lta}{\lower.5ex\hbox{\ltaraw}}
\newcommand{\gtaraw}{$\; \buildrel > \over \sim \;$}
\newcommand{\gta}{\lower.5ex\hbox{\gtaraw}}
\def\lsim{\mathrel{\rlap{\lower4pt\hbox{\hskip1pt$\sim$}}
    \raise1pt\hbox{$<$}}}                

\def\apjl{ApJ}%
\def\apjs{ApJS}%
\def\ao{Appl.~Opt.}%
\def\aap{A\&A}%
\newcommand{\TWOLPTIC}{\textsc{2LPTic}}
\newcommand{\GADGETtwo}{\textsc{GADGET-2}}
\newcommand{\Subfind}{\textsc{Subfind}}
\newcommand{\ROCKSTAR}{\textsc{ROCKSTAR}}

\newcommand{\Tiamat}{\textit{Tiamat}}

\newcommand{\TinyTiamat}{\textit{Tiny Tiamat}}
\newcommand{\MediTiamat}{\textit{Medi Tiamat}}

\title[DRAGONS II: High-z Spin and
Concentration]{\mbox{Dark-ages~Reionization~$\&$~Galaxy~Formation~Simulation~II}:
  \\Spin and concentration parameters for dark matter haloes during the Epoch of Reionization}

\author[Angel et al.]{Paul W. Angel$^{1}$, Gregory B. Poole$^{1}$, Aaron D. Ludlow$^{2}$, Alan R. Duffy$^{3}$, \newauthor Paul M. Geil$^{1}$, Simon J. Mutch$^{1}$, Andrei Mesinger$^{4}$, J. Stuart B. Wyithe$^{1}$ \\
$^{1}$ School of Physics, University of Melbourne, Parkville, Victoria, Australia: pangel@student.unimelb.edu.au\\
$^{2}$ Institute for Computational Cosmology, Dept. of Physics, University of Durham, South Road, Durham DH1 3LE, UK\\
$^{3}$ Centre for Astrophysics $\&$ Supercomputing, Swinburne University of Technology, PO Box 218, Hawthorn, VIC 3122, Australia\\
$^{4}$ Scuola Normale Superiore, Piazza dei Cavalieri 7, I-56126 Pisa, Italy\\}

\begin{document}
\date{\today}
\maketitle
\label{firstpage}
\begin{abstract}
We use high resolution N-Body simulations to study the
concentration and spin parameters of dark matter haloes in the mass range 
$10^8\, {\rm M}_{\odot}\, h^{-1} < {\rm M} < 10^{11}\, {\rm
  M}_{\odot}\, h^{-1}$ and redshifts $5{<}z{<}10$, 
corresponding to the haloes of galaxies thought to be responsible for
reionization. We build a sub-sample of equilibrium 
haloes and contrast their properties to the full population that also includes
unrelaxed systems. Concentrations are calculated by fitting both 
NFW and Einasto profiles to the spherically-averaged density profiles
of individual haloes. 
After removing haloes that are 
out-of-equilibrium, we find a $z{>}5$ 
concentration$-$mass ($c(M)$) relation that is almost flat and
well described by a simple power-law for both NFW and Einasto fits.
The intrinsic scatter around the mean
relation is $\Delta c_{\rm{vir}}{\sim1}$ (or 20 per cent) at
$z=5$. We also find that the analytic model proposed by \citet{2014MNRAS.441..378L} 
reproduces the mass and redshift-dependence
of halo concentrations. Our best-fit Einasto shape parameter, 
$\alpha$, depends on peak height, $\nu$, in a manner that is
accurately described by $\alpha {=}0.0070\nu^2{+}0.1839$. 
The distribution of the spin parameter, $\lambda$, has a weak dependence on 
equilibrium state; $\lambda$ peaks at roughly ${\sim}0.033$ for our relaxed sample, and
at ${\sim}0.04$ for the full population. The spin--virial mass relation has a 
mild negative correlation at high redshift.
\end{abstract}


\section{Introduction}
\vspace{-1mm}
In the current cosmological paradigm cold dark
matter (CDM) collapses to form gravitationally bound structures
within an expanding background universe. Known as dark matter (DM) haloes, 
these objects are initially small but undergo repeated merging to 
form ever larger systems. Galaxies form within these haloes as
in-falling gas cools and converts to stars
\citep[e.g.][]{1978MNRAS.183..341W}. Their evolution and structural 
properties therefore underpin those of the embedded galaxies. 
These ideas have evolved into the field of semi-analytic modelling 
in which galaxies are grown within an evolving 
population of dark-matter haloes extracted from purely N-Body 
simulations \citep[e.g.][see \citealt{2006RPPh...69.3101B} for a review]{2006MNRAS.365...11C, 2008MNRAS.391..481S, 2008MNRAS.388..587L}.  

The characteristics of DM haloes have been the 
subject of extensive research. Mass determines the overall size 
of the halo, but several other important parameters have also been 
identified. For example, using N-Body simulations \citep[][henceforth NFW]{1997ApJ...490..493N} 
found that the density profiles of virialised haloes can be well
described by rescaling a simple formula (hereafter known as the NFW 
profile):
\begin{equation}
\frac{\rho(r)}{\rho_{\rm{c}}} = \frac{\delta_{\rm{c}}}{(r/r_{\rm{s}})(1+r/r_{\rm{s}})^2}.
\label{eqNFW}
\end{equation}
Here $r_{\rm{s}}$ is the characteristic scale radius at
which the logarithmic density slope is equal to -2; $\delta_{\rm{c}}$
is the characteristic density contrast, and $\rho_{\rm c}=3\, H(z)^2/8\pi G$ the critical background density at redshift $z$,
where G is Newton's gravitational constant and $H$ the Hubble parameter.
These parameters can be expressed in a variety of forms. One common
approach is to use a virial mass\footnote{We define the virial mass
  of a halo as that enclosed by the radius $R_{\rm{vir}}$ within which the density is  
$\Delta = 18\pi^2 + 82(\Omega(z)-1)-39(\Omega(z)-1)^2$ 
times the background density, $\rho_{\rm{c}}$
\citep[][]{1998ApJ...495...80B}. Note that
$V_{\rm{vir}}\equiv\sqrt{G\, M_{\rm vir}/R_{\rm{vir}}}$ is 
the circular velocity at the virial radius.}, and
concentration, $c_{\rm{vir}}\equiv R_{\rm{vir}}/r_{\rm{s}}$,
defined as the ratio of the halo's virial radius to its scale
radius. 

While the NFW profile is a common description, several recent studies 
\citep[e.g.][]{2004MNRAS.349.1039N, 2008MNRAS.388....2H,2010MNRAS.402...21N} 
have shown that the density profiles of simulated haloes exhibit
small but systematic deviations from eq.~\ref{eqNFW}. The Einasto profile 
\citep[][]{1965TrAlm...5...87E}, defined
\begin{equation}
  \ln\biggl(\frac{\rho_{\rm Ein}(r)}{\rho_{-2}}\biggl)=-\frac{2}{\alpha}\biggl[\biggl(\frac{r}{r_{-2}}\biggl)^\alpha-1\biggl],
\label{eqEIN}
\end{equation}
provides a better approximation to the radial density profile 
\citep[][]{2004MNRAS.349.1039N,2013MNRAS.432.1103L}. Like the NFW
profile, eq.~\ref{eqEIN} has two scaling parameters, $r_{-2}$ and
$\rho_{-2}$, and an additional shape parameter, $\alpha$. Note that
$r_{-2}$ and $r_s$ are equivalent, and we will hereafter use the two 
interchangeably when defining the concentration of a halo. 

At low redshift the concentration parameter decreases
with increasing halo mass. NFW interpreted this finding as a result of 
hierarchical clustering: smaller haloes form earlier than more 
massive objects, when the universe was denser. They suggested that 
concentration, or equivalently the characteristic density, $\delta_c$, 
reflects the background density of the Universe at the 
halo's formation time.

The same negative trend was also seen in subsequent N-Body 
simulations \citep[e.g.][]{2001ApJ...554..114E,2002ApJ...568...52W,
  2008MNRAS.390L..64D}, and lead to the development of analytic models
attempting to explain the dependence of concentration on mass, 
redshift and cosmology. One approach relates $\delta_c$ to
the past accretion history of the halo's main progenitor. 
\cite{2002ApJ...568...52W}, for example, calculated the mass assembly
histories (MAHs) of 
simulated haloes and used a proportionality constant to relate the 
concentration to background density at the halo's time of formation,
defined by the point at which the logarithmic accretion rate falls
below a specific value. The redshift dependence of the $c(M)$ relation 
was later studied by \cite{2003ApJ...597L...9Z}, who found a weakening of the 
relation for the highest mass haloes at any redshift. By $z{\sim}$3-4 the negative trend 
is no longer present in the simulations of \cite{2008MNRAS.387..536G},
who focus on masses $M > 10^{11}{\rm M}_{\odot}/h$. The flattening 
of the $c(M)$ relation was also reported by 
\cite{2009ApJ...707..354Z} who connected halo concentrations to the 
period at which halo growth transitions from a rapid to a slow phase.
Although these models have been met with limited success overall, they provide 
a clear interpretation of why concentration depends only weakly on
mass for the most massive systems: because these haloes are forming
{\em today}, they share the same formation time, and therefore
concentration.

\cite{2007MNRAS.381.1450N} studied the $z=0$ $c(\rm{ M})$ relation
in the Millennium 
simulation \citep[][]{2005Natur.435..629S}, while 
\cite{2011ApJ...740..102K} and \cite{2012MNRAS.423.3018P} extended the
analysis to $z=6$ using both the Millennium and Bolshoi simulations.
In agreement with previous work, these authors each find a decline of 
concentration with mass. However, both \cite{2012MNRAS.423.3018P} 
and \cite{2011ApJ...740..102K} have also reported an upturn of the 
$c({\rm M})$ relation at the high mass end. On the 
other hand, Ludlow et al. (2012) demonstrated that there is no
upturn amongst relaxed haloes, and showed how the transient 
dynamical states of merging systems can result in a non-monotonic 
$c({\rm M})$ relation. While the details continue to be debated, it 
is clear overall that the diversity of halo formation 
histories play a critical role in establishing the shape and evolution
of the $c({\rm M})$ relation \citep[][]{2014MNRAS.441..378L,2015arXiv150200391C}.  

While the low redshift mass-concentration 
relation is well studied, at high $z$ the relation is poorly 
constrained. For example, \cite{2012MNRAS.423.3018P} and 
\cite{2014arXiv1407.4730D} find a high-mass upturn (above a few times
$10^{10}\, h^{-1}\, {\rm M}_\odot$) at $z=5$ amongst the full
halo population.
At similar masses, \cite{2014arXiv1402.7073D}, find a relation with a 
slightly positive slope, whereas
\cite{2015arXiv150506436H} report a weak negative slope
that flattens by $z{=}9$. These authors, however, imposed
different equilibrium cuts on their halo samples, which hampers
a direct comparison with their results.
Given this unsettled state of affairs it is clear that there is some 
debate on the precise nature of the high redshift $c(\rm{ M})$
relation and the role played by unrelaxed haloes. 

For example, a
halo suffering a merger is unlikely to have a simple, smooth density
profile, and will take time to settle back into equilibrium. This 
situation becomes increasingly important at high redshift due 
to the elevated merger rates of potentially star-forming haloes.
These sorts of concerns led \cite{2007MNRAS.381.1450N} to introduce 
three physically-motivated parameters to identify systems far 
from equilibrium: 1) the mass-fraction in substructure
$f_{\rm{sub}}={\rm M_{sub}}(< {\rm R_{vir}})/{\rm M_{vir}}$; 2) the 
offset between the halo's center of mass and its most-bound 
particle, $x_{\rm{off}}$, and 3) the pseudo-virial-ratio of kinetic
and potential energies, $\varphi$. The effectiveness of these
parameters in isolating relaxed DM haloes is further discussed in 
\cite{2008MNRAS.387..536G} and \cite{2012MNRAS.427.1322L}.

A detailed study of these parameters with regard to dynamical 
relaxation at high redshift is provided by the first paper of the DRAGONS\footnote{The Dark-ages, Reionization And Galaxy-formation Observables Numerical Simulation project;  \url{ http://dragons.ph.unimelb.edu.au/} } series, Poole et al. (2015b) (hereafter referred to 
as Paper I), who examine the behaviour of these equilibrium 
diagnostics during the relaxation phase that follows a significant 
merger or accretion event during the Epoch of Reionization. Their results suggest that, across the mass range of our 
simulations, $10^{8} < {\rm M}/[h^{-1}\, {\rm M}_\odot] < 10^{11}$, 
and for $z>5$, standard relaxation values for $f_{\rm{sub}}$, $x_{\rm{off}}$ and $\varphi$ obtained from low redshift studies are very effective at identifying systems relaxing from halo formation or recent mergers at high redshift. 

Concentration is not the only relevant halo property
for galaxy formation. Halo spin also plays an important role 
in semi-analytic models, since angular momentum conservation 
determines the size of galactic disks
\citep[e.g.][]{1998MNRAS.295..319M,2011MNRAS.413..101G}, 
which in turn determine their star formation rates
\citep[][]{1959ApJ...129..243S, 1998ApJ...498..541K}. 
A halo's angular momentum is often expressed as a dimensionless 
spin parameter:
\begin{equation}
\lambda = \frac{J_{\rm{vir}}}{\sqrt{2}M_{\rm{vir}}V_{\rm{vir}}R_{\rm{vir}}},
\end{equation}
where $J_{\rm{vir}}$ is the total angular momentum within ${\rm R_{vir}}$. 

 Most studies of the spin parameter have focused on the 
distribution of spins and its dependence on halo mass
\citep[e.g.][]{2007MNRAS.376..215B,2007MNRAS.381.1450N,2008ApJ...678..621K,2010crf..work...16M}. 
At any redshift, halo spins are distributed approximately log-normally,
and peak at $\lambda_0 {\sim} 0.03{-}0.04$.
At low redshift, spins are approximately independent of mass
but gain a slight negative correlation at higher redshifts 
\citep[][]{2008ApJ...678..621K,2010crf..work...16M}. 
Recently, \cite{2014arXiv1402.7073D} measured the redshift evolution 
of the $\lambda - M_{\rm{vir}}$ relation, reporting a weak negative
correlation at $z=5$.

In this work we use the {\em Tiamat}
simulation suite to extend the study of concentration and spin to 
redshifts $z>5$. Our simulations were designed to resolve halo masses 
relevant for galaxy formation during this high-redshift epoch. The
purpose of our study is to measure the structural and
dynamical properties of haloes that are necessary for forthcoming 
semi-analytic models of reionization.

We organise the paper as follows. In Section 2 we describe
the numerical simulations, including halo finding, analysis 
techniques, and our parametrization of concentration and spin. 
In Section 3 we present our concentration--mass relation and its 
redshift dependence, and in Section 4 the spin distribution, and 
its mass and redshift dependence. Finally, in Section 5 we 
summarise our main results.

\section{Numerical Simulation}

Our analysis focuses on DM haloes identified in three cosmological
N-body simulations. These include a 2160$^3$-particle, 67.8 Mpc$/h$ cubed 
box (the $\Tiamat$ simulation) and two 
smaller but higher resolution volumes of 10 and 22.6 (Mpc$/h$)$^3$. 
Each run was carried out with $\GADGETtwo$ \citep[][]{2001NewA....6...79S, 2005MNRAS.364.1105S} 
with RAM (random-access memory) consumption changes in accordance with 
those detailed in \citet[][]{2015MNRAS.449.1454P}. For each run, the
Plummer-equivalent softening length was $1/50^{\rm{th}}$ of the mean 
Lagrangian inter-particle spacing, and the integration accuracy
parameter, $\eta$, is set to 0.025, as motivated by the convergence 
study presented in \citet[][]{2015MNRAS.449.1454P}. Initial conditions were 
generated using 2nd order perturbation theory \citep[using the code $\TWOLPTIC$\footnote{Code
  was obtained from http://cosmo.nyu.edu/roman/2LPT/ with alterations
  to allow for greater than 1625$^3$ particles.},][]{2006MNRAS.373..369C} 
at $z=99$ and each simulation was run down to $z=5$; 100 snapshots of 
particle data were taken equally spaced in time from $z=35$ to $z=5$ 
(one every 11 Myr). Cosmological parameters for each box were chosen 
to be consistent with the Planck 2015 data release
\citep[][]{2015arXiv150201589P} 
$(h, \Omega_{\rm{m}}, \Omega_{\rm{b}}, \Omega_\Lambda, \sigma_8,
n_{\rm{s}})$ 
= $(0.678, 0.308, 0.0484, 0.692, 0.815, 0.968)$. The relevant
numerical parameters are summarised in Table \ref{Simulation
  Parameters}. A more detailed discussion of these simulations can be 
found in Paper I.

\begin{table}

\begin{center}
\begin{tabular}{|l|p{0.6cm} p{1.0cm} p{1.2cm} p{1.0cm} |}
\hline
Simulation  & $N_{\rm{p}}$  & L [Mpc$/h$] & $m_{\rm{p}}$ [$\rm{M}_{\odot}/h$] & $\epsilon$ [kpc$/h$] \\
\hline
\Tiamat     & $2160^3$ & 67.8              & $2.64 {\times} 10^6$ & 0.63   \\
\MediTiamat & $1080^3$ & 22.6              & $7.83 {\times} 10^5$ & 0.42    \\
\TinyTiamat & $1080^3$ & 10.0              & $6.79 {\times} 10^4$ & 0.19    \\ 
\hline
\end{tabular}
\caption{Summary of simulation parameters. $N_{\rm{p}}$ is the total number of particles, L is the length of the box, $m_{\rm{p}}$ is the mass of each particle and $\epsilon$ is the gravitational force softening length.}
\label{Simulation Parameters}
\end{center}

\end{table}

\begin{figure}
\includegraphics[scale=0.4]{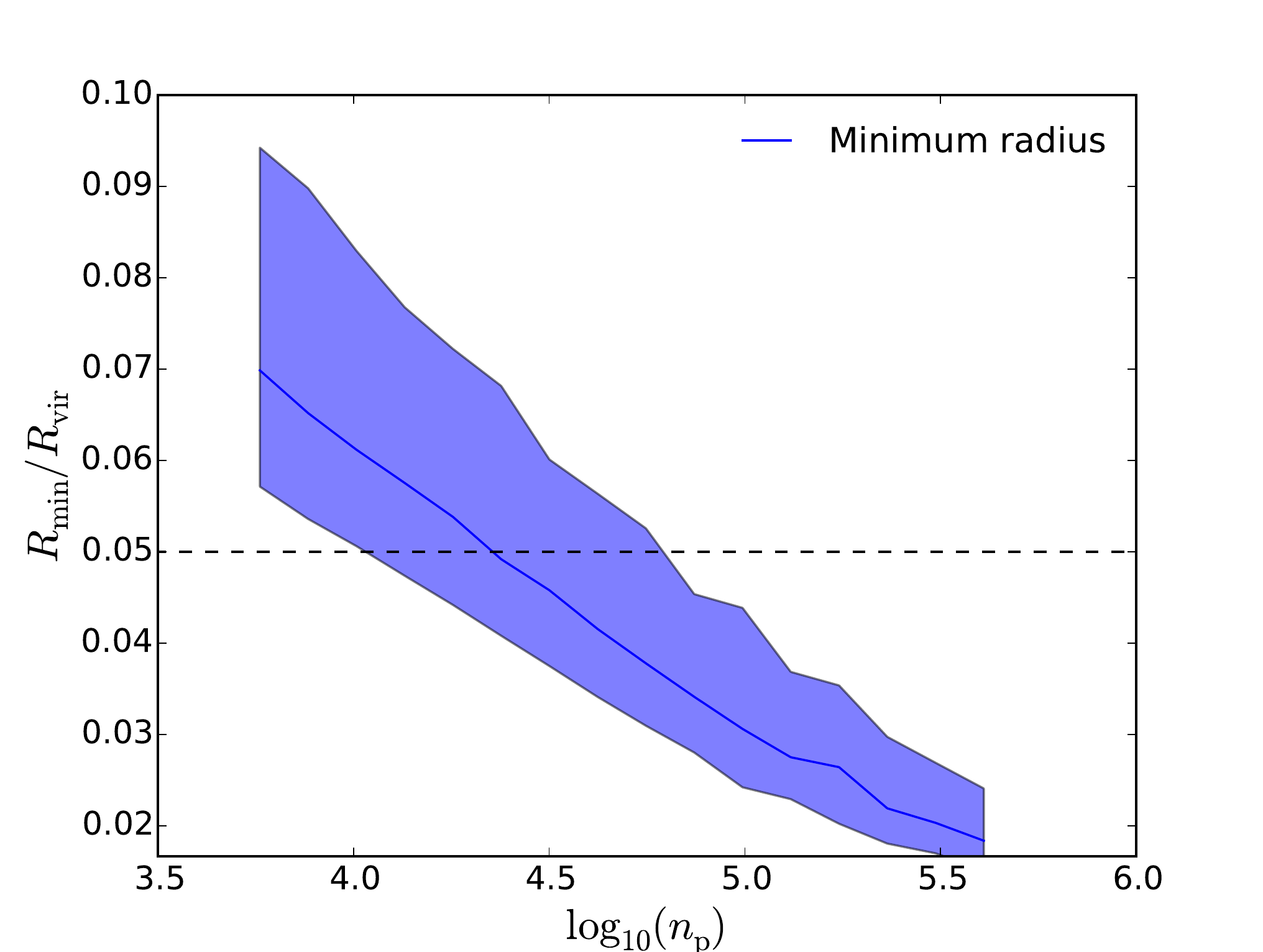}
\caption{The ratio between the minimum bin size of the halo profile in
  dark blue and the size of the virial radius, plotted as a function
  of particle number for the $\Tiamat$ simulation. The solid line is the median while the shaded area
  is the scatter (68 per cent confidence interval). The largest bin size of $0.09R_{\rm{vir}}$ indicates our bins begin well inside the halo scale radius for the halo masses considered here.}
\vspace{-2mm}
\label{Minimum Bin Radius}
\end{figure}

\subsection{Halo Finding}
Haloes were identified in each simulation snapshot using $\Subfind$
\citep[][]{2001MNRAS.328..726S}. This produces two outputs: the 
first contains structures found by a friends-of-friends (FoF) 
algorithm (we adopt a linking length of 0.2 times the mean inter 
particle spacing); the second is obtained by dissecting each FoF
group into its self-bound ``substructure''. This results is 
a central ``main halo'', typically containing $>90$ per cent of its 
virial mass, and a group of lower-mass subhaloes which trace
the undigested cores of previous merger events.

Each main halo and its substructures were further analyzed to catalog their 
basic properties and to build their spherically averaged profiles. 
A (FoF or substructure) halo was required to have a at least 32
particles to be included, resulting in
a minimum halo mass of $8.4\times10^7 \rm{M}_{\odot}/h$ in $\Tiamat$, 
$2.5\times10^7 \rm{M}_{\odot}/h$ in $\MediTiamat$ and 
$2.2\times10^6 \rm{M}_{\odot}/h$ in $\TinyTiamat$. However, when 
estimating concentrations, a stricter limit 
of $n_{\rm{p}}{>}5000$ within the bound structure was imposed to 
ensure that the halo's inner regions are well-resolved. For the 
purpose of estimating spin parameters, this limit is relaxed to 
600 particles. These parameter choices are further discussed and motivated 
in Section 2.3. 

\subsection{Density Profiles and Concentration Estimates}
For each main halo, spherically-averaged density profiles were constructed in bins
containing an equal number of particles. Only particles considered 
by $\Subfind$ to be bound to the central haloes were used. 
The number of bins was increased along with the number of halo particles.
We imposed a minimum of 5 bins for the smallest 
haloes, and a maximum of 1000 bins for the largest (reached as $n_{\rm{p}}$ tends 
to $10^6$). For example, haloes with $n_{\rm{p}} {=}$ 5000 
have 25 bins, rising to ${\sim}$125 bins for haloes containing
$10^5$ particles. 

Best-fit NFW and Einasto profiles were obtained by minimizing
\begin{equation}
\psi = \frac{1}{N} \sum\limits_{\rm{i}}^{\rm{N}} \frac{\Delta r_{\rm{i}}}{r_{\rm{i}}}(\log \rho_{\rm{i}} - \log \rho_{\rm{NFW(Ein)}})^2,
\end{equation}
where N is the number of bins, $\rho_{\rm{i}}$ is the
density in bin $i$, and $\rho_{\rm{NFW(Ein)}}$ is the corresponding density of the 
NFW (Einasto) model. The factor
$\Delta r_{\rm{i}} / r_{\rm{i}}$ (the width of bin $i$ divided by the 
median radius of particles in it) appropriately weights bins of differing size. Note
that only bins in the radial range $0.05R_{\rm{vir}} < r_{i} < 0.8R_{\rm{vir}}$ 
were used \citep[we have verified that the minimum bin radius is always larger than the
convergence radius defined in][]{2003MNRAS.338...14P}.  

In Figure \ref{Minimum Bin Radius} we show the 
ratio of the minimum bin size to the halo virial radius. For the 
smallest resolved haloes ($n_{\rm{p}} {=}$ 5000) this ratio falls to 
${\sim}$ 12, and thus scale radii corresponding to concentrations as 
high as 12 could be resolved even for the smallest haloes considered. 
We find concentrations are almost always ${<}$10. This indicates that 
our bins span a sufficient portion of the halo profile to provide 
reliable estimates of $r_{\rm{s}}$.

\subsection{Resolution and Virialisation Cuts}

In defining our sample of equilibrium DM haloes we impose 
both dynamical and resolution criteria following the procedure established 
by \cite{2007MNRAS.381.1450N}, \cite{2008MNRAS.387..536G} and 
\cite{2012MNRAS.427.1322L}. The behaviour of these equilibrium
diagnostics over the mass and redshift 
range probed by the $\Tiamat$ simulation suite is discussed further in Paper
I. The criteria defining relaxed haloes include upper limits on the following
three quantities: \renewcommand{\theenumi}{\Roman{enumi}}%

\begin{enumerate} 
\item[I)] The fraction of mass found in satellite subhaloes, 
$f_{\rm{sub}} = M_{\rm{sub}}(<R_{\rm{vir}})/M_{\rm{vir}} < 0.1$. As discussed 
in \cite{2007MNRAS.381.1450N} a high fraction of mass in substructure
may be indicative of a recent merger. Such a halo 
will not, in general, have a smooth, spherically-averaged density profile. 

\item[II)] The offset between the position of the most-bound-particle, 
$\vec{r}_{\rm{mbp}}$, and center-of-mass, $\vec{r}_{\rm{com}}$, in
units of virial radius: $x_{\rm{off}} = |\vec{r}_{\rm{mbp}} -\vec{r}_{\rm{com}}|/R_{\rm{vir}} < 0.07$. 
This is complementary to $f_{\rm{sub}}$ as it includes mass from
unresolved subhaloes, or merging haloes that have their center 
just outside the virial radius and consequently are not included 
in $f_{\rm{sub}}$. 

\item[III)] The pseudo-virial ratio of kinetic and potential 
energies, $\varphi = 2K/|U| < 1.35$. This criterion tends to be sensitive to 
haloes at the pericenter of a merger that may not be flagged
by the above two parameters. 
\end{enumerate}

\cite{2007MNRAS.381.1450N} find that these restrictions, although arbitrary 
in value, provide a simple and physically motivated method to exclude 
haloes that are not well described by an NFW profile. In Paper I these
parameters were also shown to be discriminate between haloes that 
have either recently doubled in mass, or suffered a major or minor merger, within the last 
$1{-}2$ dynamical times. We therefore adopt them as our standard 
equilibrium criteria, and use them to split our full halo sample into 
a relaxed sub-sample, which we analyse separately. 

A further criterion is imposed to select only well-resolved 
haloes from both the relaxed and unrelaxed populations, in order to 
fit a reliable radial halo profile. For the halo concentration 
analysis a lower limit on particle number ($n_{\rm{p}}$) of 
$n_{\rm{p}}{>}5000$ is imposed. This is derived from work studying 
convergence of NFW and Einasto profile fits in
\cite{2007MNRAS.381.1450N} and \cite{2008MNRAS.387..536G}. This 
ensures that the inner portion of the halo is well enough resolved 
to measure the scale radius $r_{\rm{s}}$. We impose a restriction 
of $n_{\rm{p}}>600$ particles for the spin parameter measurement 
following \cite{2008ApJ...678..621K}, who obtain the limit by 
comparing the measured energy from Monte Carlo realisations of 
analytic NFW profiles. 

The dynamics of hierarchical growth means that at the high 
redshifts studied here, many of our haloes will be far from 
equilibrium and thus have ill-defined values for concentration and 
spin. These sample cuts are designed to remove such systems 
and to keep transients out of our analysis as much as possible. 
For example, a halo which has just 
undergone a major merger may be comprised of two large, high 
density clumps, and consequently have a high $x_{\rm{off}}$ and 
poorly defined center. The density profile of such a system cannot
be captured by simple spherical averages, and structural properties
estimated from NFW or Einasto fits are meaningless. We stress, however, that there is a 
continuum of values for $x_{\rm{off}}$, $f_{\rm{sub}}$ and $\varphi$;
the particular values chosen to separate relaxed haloes from unrelaxed
are the result of extensive past investigations 
\citep[see, e.g.,][and Paper I of this series]{2007MNRAS.381.1450N,2012MNRAS.427.1322L}.
Our choices of resolution and dynamical classification also
simplify comparison with previous studies.

To quantify the effects of our sample selection, we note that in 
$\Tiamat$ there are 14391 haloes with more than 5000 
particles at redshift $z\sim5$, but only 4433 (or $\sim$30 per cent) of 
these satisfy our relaxation criteria; this reduces to $\sim$15 per cent 
by $z\sim10$. These numbers underline the importance of the dynamical 
state of haloes at high redshift. Nevertheless parametrised fits to
the entire population are useful for many semi-analytic calculations 
\citep[e.g.][]{2014MNRAS.439.2728M,2015MNRAS.451.2840S} and for this 
reason we report fits to both our full halo sample as well as the
relaxed sub-sample. Further details regarding the dynamical state of 
high redshift haloes are presented in Paper I.

\begin{figure*}
\begin{minipage}{170mm}
\begin{center}
\includegraphics[width=170mm]{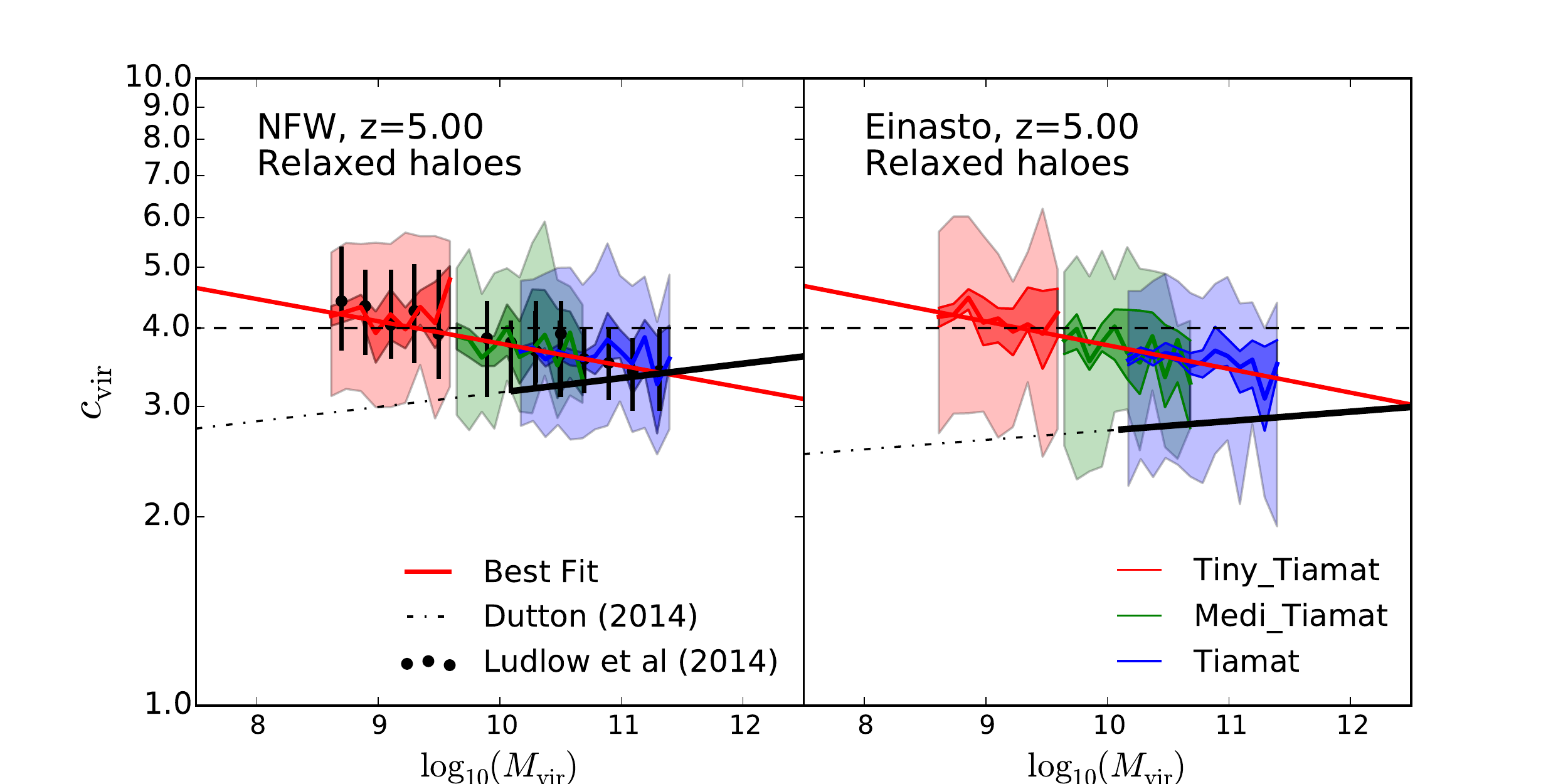}  
\caption{Concentration--mass relation of relaxed central haloes at
  $z{=}5$. Left and Right panels are the NFW and Einasto
  concentrations respectively. Inner shaded region denotes the bootstrapped 90 per cent
  confidence interval on the median. The outer shaded region shows the
  $68$ per cent scatter. The line of best fit is fit to the median using the
  Monte Carlo Markov Chain method implemented in the Python package
  \textsc{emcee} \protect\citep[][]{2013PASP..125..306F}. This produces
  $c_{\rm{vir}}=3.8\pm0.4 (\rm{M}/[10^{10}\rm{M}_{\odot}h^{-1}])^{-0.035\pm0.005}$
  at $z{=}5$ for NFW fits and
  $c_{\rm{vir}}=3.8\pm0.4 (\rm{M}/[10^{10}\rm{M}_{\odot}h^{-1}])^{-0.039\pm0.005}$
  for Einasto fits. The fits from \protect\cite{2014arXiv1402.7073D} are also
  shown with thick solid lines representing fitting formula over their
  derived mass range, although we emphasise \protect\cite{2014arXiv1402.7073D}
  employ different relaxation and resolution criteria for their
  sample. Thin extensions of these lines represent these fits
  extrapolated to lower masses. The $\Tiamat$ simulation is designed
  to study reionization and explores a mass range below previous
  studies. The dashed black line at $c_{\rm{vir}}=4$ is added for a
  point of reference. Solid black points represent the
  \protect\cite{2014MNRAS.441..378L} model where halo concentrations are
  calculated from the median accretion history in the mass bin. Error
  bars are derived from the 68 per cent scatter in accretion histories.}
\label{Resolved Concentration z 5}
\end{center}
\end{minipage}
\end{figure*}

\begin{figure*}
\begin{minipage}{170mm}
\begin{center}
\includegraphics[width=170mm]{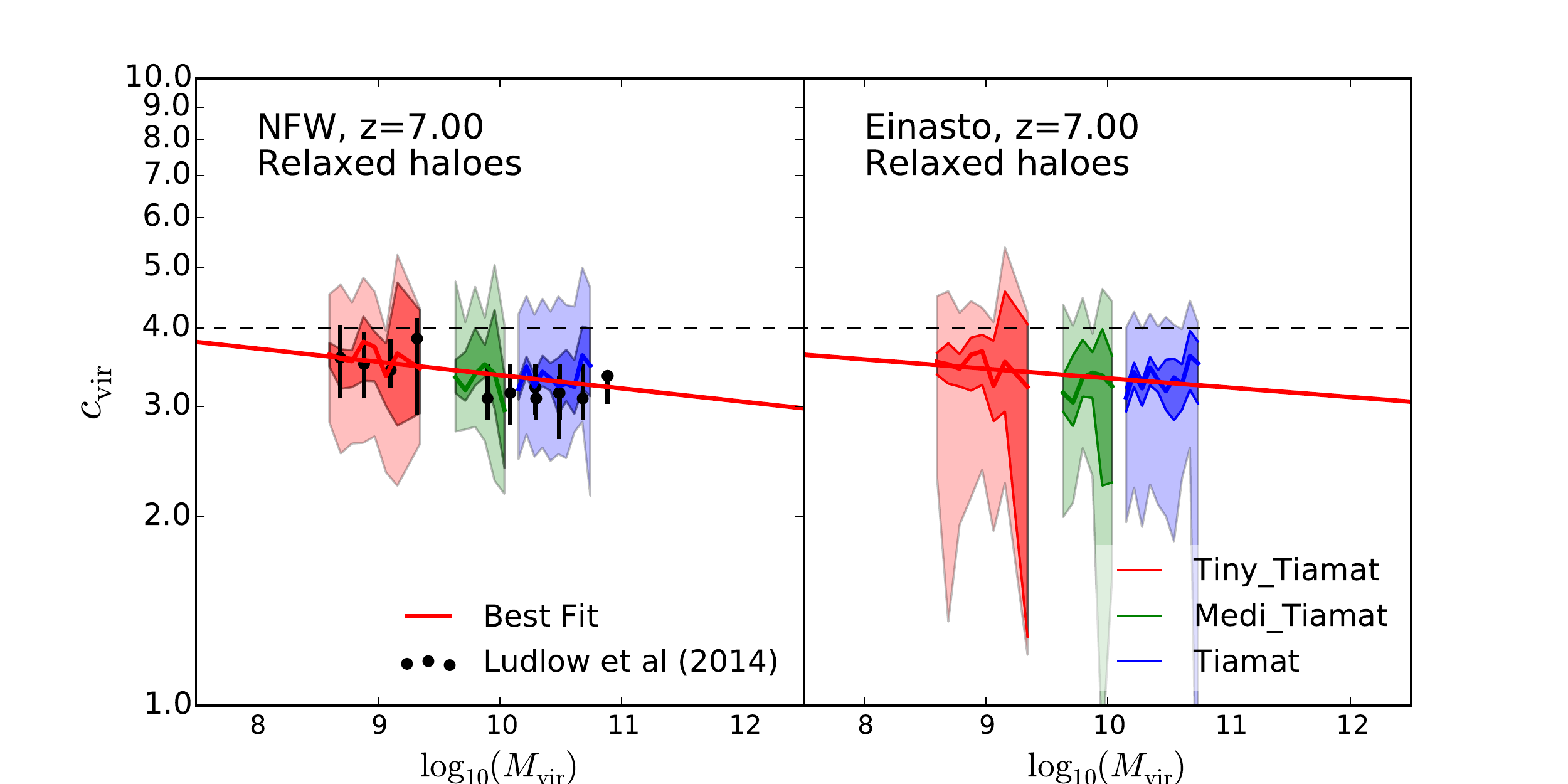}  
\caption{The same concentration mass relations for relaxed central
  haloes as shown in Figure \ref{Resolved Concentration z 5}, but now
  at $z{=}7$. Inner shaded region denotes the bootstrapped 90 per cent
  confidence interval on the median. The outer shaded region shows the
  $68$ per cent scatter. The line is fit to the median and gives
  $c_{\rm{vir}}$ =
  $3.4\pm0.6(\rm{M}/[10^{10}\rm{M}_{\odot}h^{-1}])^{-0.019\pm0.008}$ at
  $z{=}7$ for NFW and
  $3.3\pm0.6(\rm{M}/[10^{10}\rm{M}_{\odot}h^{-1}])^{-0.018\pm0.008}$ for
  Einasto profiles. The dashed black line at $c_{\rm{vir}}=4$ is added
  for a point of reference. Solid black points represent the
  \protect\cite{2014MNRAS.441..378L} model where halo concentrations are
  calculated from the median accretion history in the mass bin. Error
  bars are derived from the $68$ per cent scatter of the accretion histories.}
\label{Resolved Concentration z 7}
\end{center}
\end{minipage}
\end{figure*}

\section{Concentration Mass Relation}

\subsection{The concentration mass relation for relaxed haloes}

The $c(M)$ relations for our equilibrium haloes at $z{=}5$ and $z{=}7$
are shown in Figures \ref{Resolved Concentration z 5} 
and \ref{Resolved Concentration z 7}, respectively. Both NFW and Einasto 
concentrations are plotted. The inner shaded region shows the 
bootstrapped 90 per cent confidence interval on the median for mass bins 
containing at least 20 haloes. The outer shaded area fills the 68 per cent scatter
in individual concentration estimates. 
We find a weak trend of decreasing concentration with mass at $z{=}5$ 
for both NFW and Einasto fits. This trend becomes shallower as redshift 
increases. By redshift 9 there is no trend apparent for either set of fits (see Table \ref{Relaxed concentration fit values}).
The $c({\rm M})$ relations obtained from both NFW and Einasto
fits are similar over this mass range. We note that the systematic 
difference between NFW and Einasto concentrations is $\Delta c_{\rm{vir}}<$0.1,
which is smaller than the 
change in concentration from the lowest to highest masses studied here. 

Best-fitting power laws at $z{=}5$ are
$3.8\pm0.4 \, ({\rm M}/[10^{10}\, h^{-1}\, {\rm M}_{\odot}])^{-0.039\pm0.005}$ for NFW fits and
$3.8\pm0.4 \, ({\rm M}/[10^{10}\, h^{-1}\, {\rm M}_{\odot}])^{-0.039\pm0.005}$ 
for Einasto fits. Best-fit parameters are obtained using the 
Monte Carlo Markov Chain (MCMC) method implemented with the 
\textsc{emcee} package \citep[][]{2013PASP..125..306F}. The quoted 
errors are the $68$ per cent confidence interval derived from the posterior distribution. We 
find intrinsic scatter in the $c({\rm M})$ relation 
of $\Delta c_{\rm{vir}}\sim1.0$ (or 20 per cent) for fits to both NFW and Einasto profiles.
Best-fit power-law relations (for both NFW and Einasto fits) 
are provided in Table \ref{Relaxed concentration fit values} for a
range of redshifts. 

\begin{figure}
\includegraphics[scale=0.4]{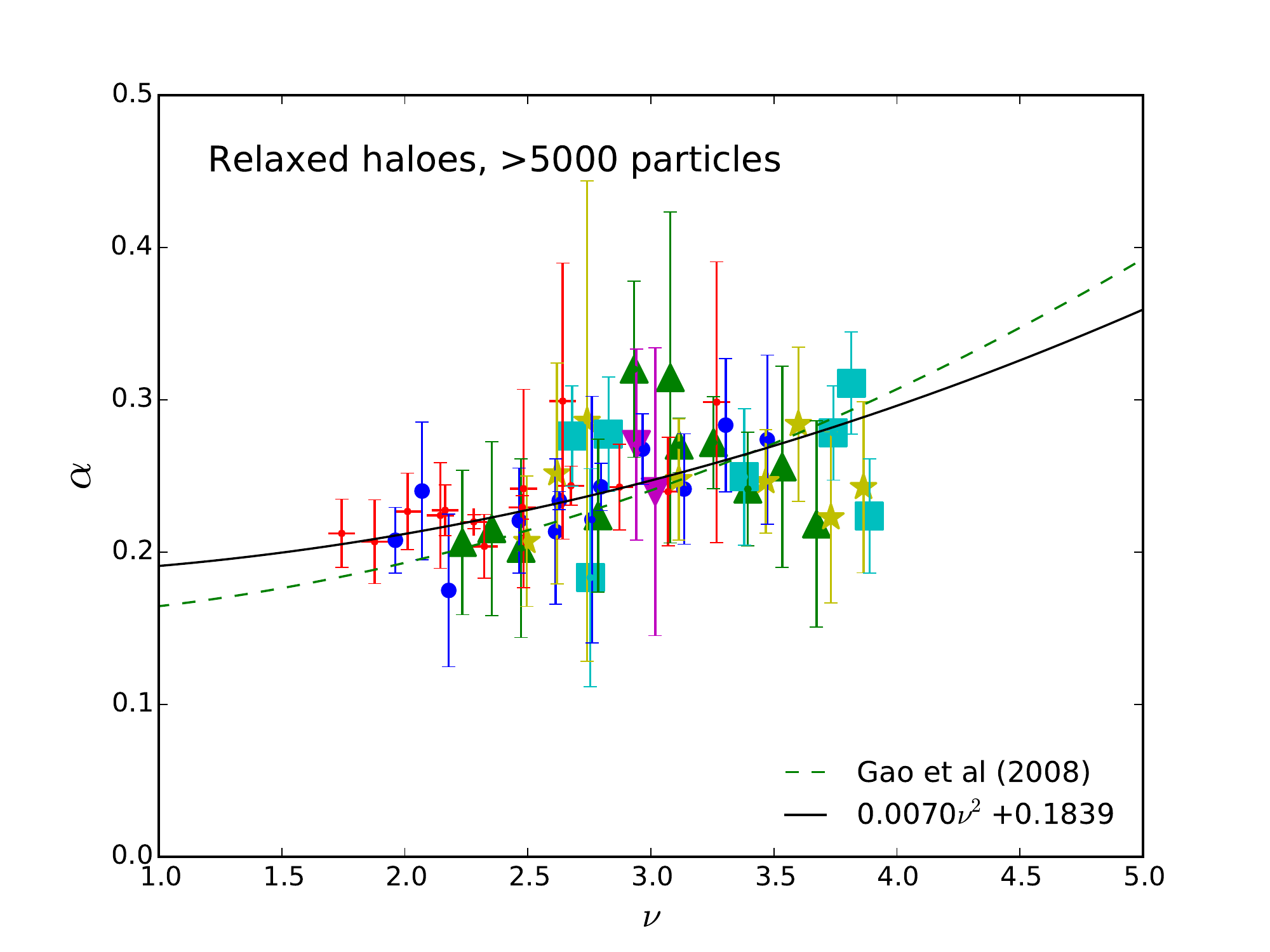}  
\caption{The Einasto profile $\alpha-\nu$ relation. At each redshift
  the median $\alpha$ is plotted for bins containing $>20$ haloes,
  with the errors derived from the bootstrapped $90$ per cent confidence interval on the median. Each symbol denotes a different redshift; red pluses denote $z{=}5$, blue filled circles denote $z{=}6$, green triangles denote $z{=}7$, yellow stars denote $z{=}8$, cyan squares denote $z{=}9$ and magenta inverted (point down) triangles denote $z{=}10$.  We find a similar $\alpha-\nu$ relation to the low redshift results of previous authors despite the redshift range of our simulations being $z>5$.}
\label{Alpha nu}
\end{figure}

In addition to the two scaling variables, the Einasto profile has a
shape parameter, $\alpha$. Previous studies have found that $\alpha$
depends in a complex way on both halo mass and redshift, but follows
a simple relation when expressed in terms of the dimensionless ``peak
height'' mass parameter, $\nu=\delta_{\rm sc}/\sigma({\rm M},z)$. Here $\delta_{\rm sc}=1.686$ is the density threshold for the
collapse of a spherical top-hat density perturbation, and $\sigma({\rm M},z)$
is the rms density fluctuation in spheres enclosing mass ${\rm M}$.
Figure \ref{Alpha nu}
shows the $\alpha - \nu$ relation obtained from our Einasto fits. 
We find a similar $\alpha-\nu$ relation to the previous authors --
the fit from \citet[][]{2008MNRAS.387..536G} is shown as a dashed
line --  although 
our simulations cover a different mass and redshift range. Our
best fit for the $\alpha-\nu$ relation is $\alpha = 0.007\nu^2+0.1839$.

In order to check for any subtle bias in our fits we have also
constructed $c({\rm M})$ and $\alpha(\nu)$ relations using the 
median density profiles obtained by stacking haloes in narrow 
mass bins. This smooths out any unique features of individual systems
and allows for a robust estimate of the median structural properties
of haloes of a given mass. For our relaxed population we recover the $c({\rm M})$
to within $\Delta c {\sim}0.1$, for both 
NFW and Einasto fits. We choose to use the individual fits 
when computing the best fit $c({\rm M})$ relation. 

The weak trend in concentration with mass 
found in our simulation is in qualitative agreement with previous 
work that found a negative trend at low redshift that becomes progressively
shallower with increasing $z$. For example, both the shallow negative 
slope and the magnitude of our Einasto concentrations 
are in good agreement with \cite{2015arXiv150506436H}. 

In Figure \ref{Resolved Concentration z 5} 
we also plot the $c({\rm M})$ relations from \cite{2014arXiv1402.7073D} 
(hereafter DM14), also obtained from both NFW and Einasto fits. 
The mass range covered by their
simulations is shown as the thick solid line, while the thin 
dot-dashed line is an extrapolation.
As DM14 employ different relaxation and resolution criteria, 
the differences we observe, although slight, are unsurprising. However, the shape of 
our trend at redshift 5 is in qualitative disagreement with DM14, 
who find that a positive trend emerges at $z{=}5$ for both Einasto 
and NFW concentrations. We also find a higher normalization (about 25
per cent) than DM14 at ${\sim}10^{10}M_{\odot}h^{-1}$. 

The method of our analysis differs in a few
significant ways from DM14. We do not speculate on the exact combination of these 
differences that effects the $c({\rm M})$ relation but note that: firstly, halo profiles in DM14 are fit out to 1.2$R_{\rm{vir}}$ while $\Tiamat$ profiles are only fit out to 
0.8$R_{\rm{vir}}$, and second, different resolution and relaxation
criteria were used. DM14 adopt a minimum halo mass corresponding 
to 500 particles, and define relaxed haloes as those satisfying 
$x_{\rm{off}} < 0.07$ and $\rho_{\rm{rms}} < 0.5$, where 
\begin{equation}
\rho_{\rm{rms}} = \sqrt{\frac{1}{\rm{N}}\sum^{\rm{N}}_{i}(\ln\rho_{\rm{i}} - \ln\rho_{\rm{\rm{Fit}}})^2}
\end{equation}
is the rms deviation between the haloes density profile and 
the best-NFW fit. 

However, we do find good agreement between our 
$c({\rm M})$ relation results and the model proposed 
by \cite{2014MNRAS.441..378L}, as indicated by the black dots in
Figures \ref{Resolved Concentration z 5} and \ref{Resolved
  Concentration z 7}. The black dots are obtained by finding the
median scale radius such that the mean density within this radius is 
equal to ${\sim}725$ times the critical density when its progenitor 
first achieved this mass, i.e. $<\rho>=725\times \rho_{\rm{crit}}$ 
\citep[][]{2013MNRAS.432.1103L}. In \cite{2013MNRAS.432.1103L} this relation is actually $<\rho>=776\times \rho_{\rm{crit}}$, as it is calibrated using a different cosmology, and to the $M_{200}$ defintion of mass.
We adjust this relation to $<\rho>=725\times \rho_{\rm{crit}}$ to 
allow for our use of the $M_{\rm{vir}}$ mass definition and the different cosmology of $\Tiamat$.
This reproduces our median $c({\rm M})$ relations. We emphasise there is no fit to the 
halo density profiles here, only the accretion histories are
required.  

\renewcommand{\arraystretch}{1.2}

\begin{table*}
\begin{minipage}{170mm}
\begin{center}
\begin{tabular}{| l | c  c  c |}
\hline
$z$ & Relaxed NFW Best Fit & Relaxed Einasto Best Fit & $N_{\rm{sample}}$ \\
\hline
5 & $(3.8\pm0.4)(\frac{M}{10^{10}M_{\odot}h^{-1}})^{(-0.035\pm0.005)}$ & $(3.8\pm0.4)(\frac{M}{10^{10}M_{\odot}h^{-1}})^{(-0.039\pm0.005)}$ & 4443,738,864\\ 
6 & $(3.6\pm0.5)(\frac{M}{10^{10}M_{\odot}h^{-1}})^{(-0.023\pm0.006)}$ & $(3.5\pm0.5)(\frac{M}{10^{10}M_{\odot}h^{-1}})^{(-0.022\pm0.007)}$ & 2303,412,652 \\
7 & $(3.4\pm0.6)(\frac{M}{10^{10}M_{\odot}h^{-1}})^{(-0.019\pm0.008)}$ & $(3.3\pm0.6)(\frac{M}{10^{10}M_{\odot}h^{-1}})^{(-0.018\pm0.008)}$ & 1043,221,417 \\
8 & $(3.3\pm1.0)(\frac{M}{10^{10}M_{\odot}h^{-1}})^{(0.005\pm0.013)}$  & $(3.3\pm1.2)(\frac{M}{10^{10}M_{\odot}h^{-1}})^{(0.013\pm0.016)}$ & 391,109,267 \\
9 & $(3.2\pm1.2)(\frac{M}{10^{10}M_{\odot}h^{-1}})^{(0.003\pm0.016)}$  & $(3.2\pm1.5)(\frac{M}{10^{10}M_{\odot}h^{-1}})^{(0.003\pm0.021)}$ & 134,49,155 \\
\hline
\end{tabular}
\caption{Best fit values from the relaxed population for NFW-derived and
  Einasto-derived $c({\rm M})$ relations
  $c_{\rm{vir}}$ =
  $A(\rm{M_{\rm{vir}}}/[10^{10}M_{\odot}h^{-1}])^B$. $N_{\rm{sample}}$
  denotes the number of haloes in the sample for $\Tiamat,
  \MediTiamat, \TinyTiamat$. Fits and errors are the median and 68 per
  cent confidence interval using the MCMC package quoted previously.}
\label{Relaxed concentration fit values}
\end{center}
\end{minipage}
\end{table*}

\begin{figure*}
\begin{minipage}{170mm}
\begin{center}
\includegraphics[width=170mm]{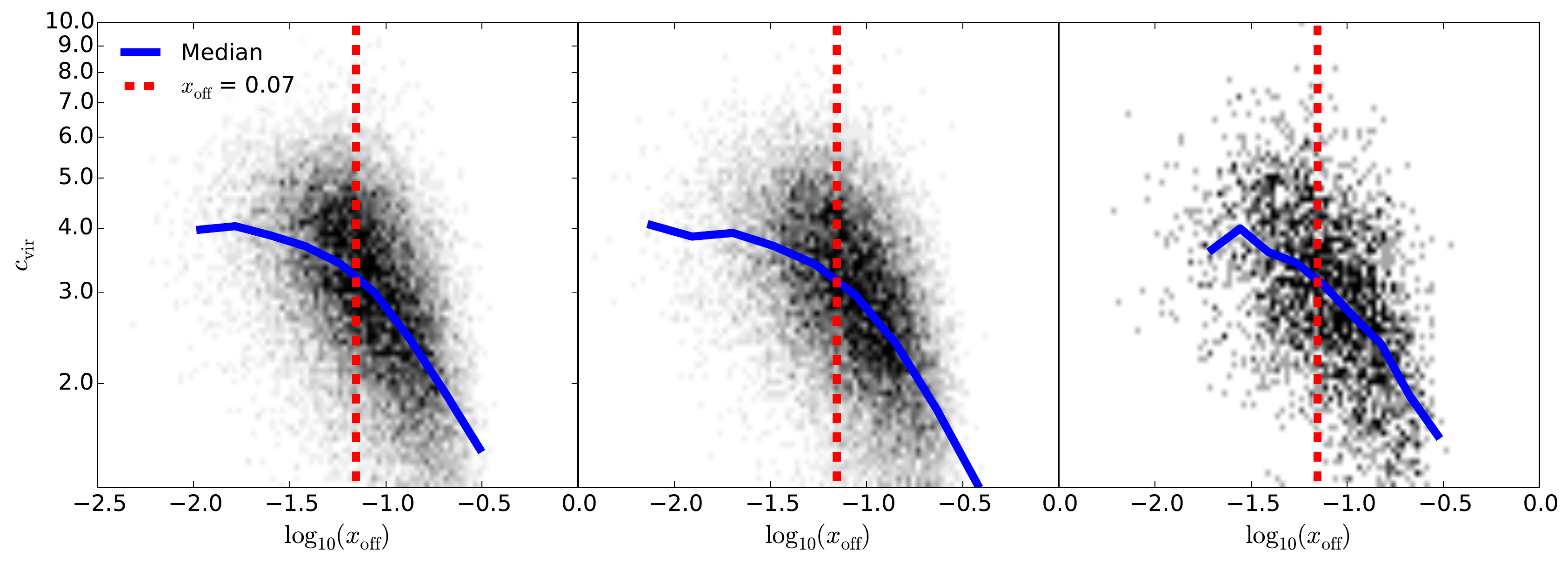} 
\caption{The dependence of concentration on $x_{\rm{off}}$ for central subgroups in several mass ranges. From left to right the ranges are [$10^{10.0}\rm{M}_{\odot}/h < M < 10^{10.5}\rm{M}_{\odot}/h$], [$10^{10.5}\rm{M}_{\odot}/h < M < 10^{11.0}\rm{M}_{\odot}/h$] and [$10^{11.0}\rm{M}_{\odot}/h < M < 10^{11.5}\rm{M}_{\odot}/h$]. Blue lines represent the median. The dashed red line denotes the relaxation criteria cut above which a halo is considered to be out of equilibrium \protect\citep[][Paper I]{2007MNRAS.381.1450N, 2008MNRAS.387..536G,2012MNRAS.427.1322L}. It can be seen that lower $x_{\rm{off}}$ parameters (i.e. more relaxed haloes) correlate with higher $c_{\rm{vir}}$ for all mass ranges.}
\label{Concentration vs xoff}
\end{center}
\end{minipage}
\end{figure*}
\vspace{2mm}

\begin{figure*}
\begin{minipage}{170mm}
\begin{center}
\includegraphics[width=170mm]{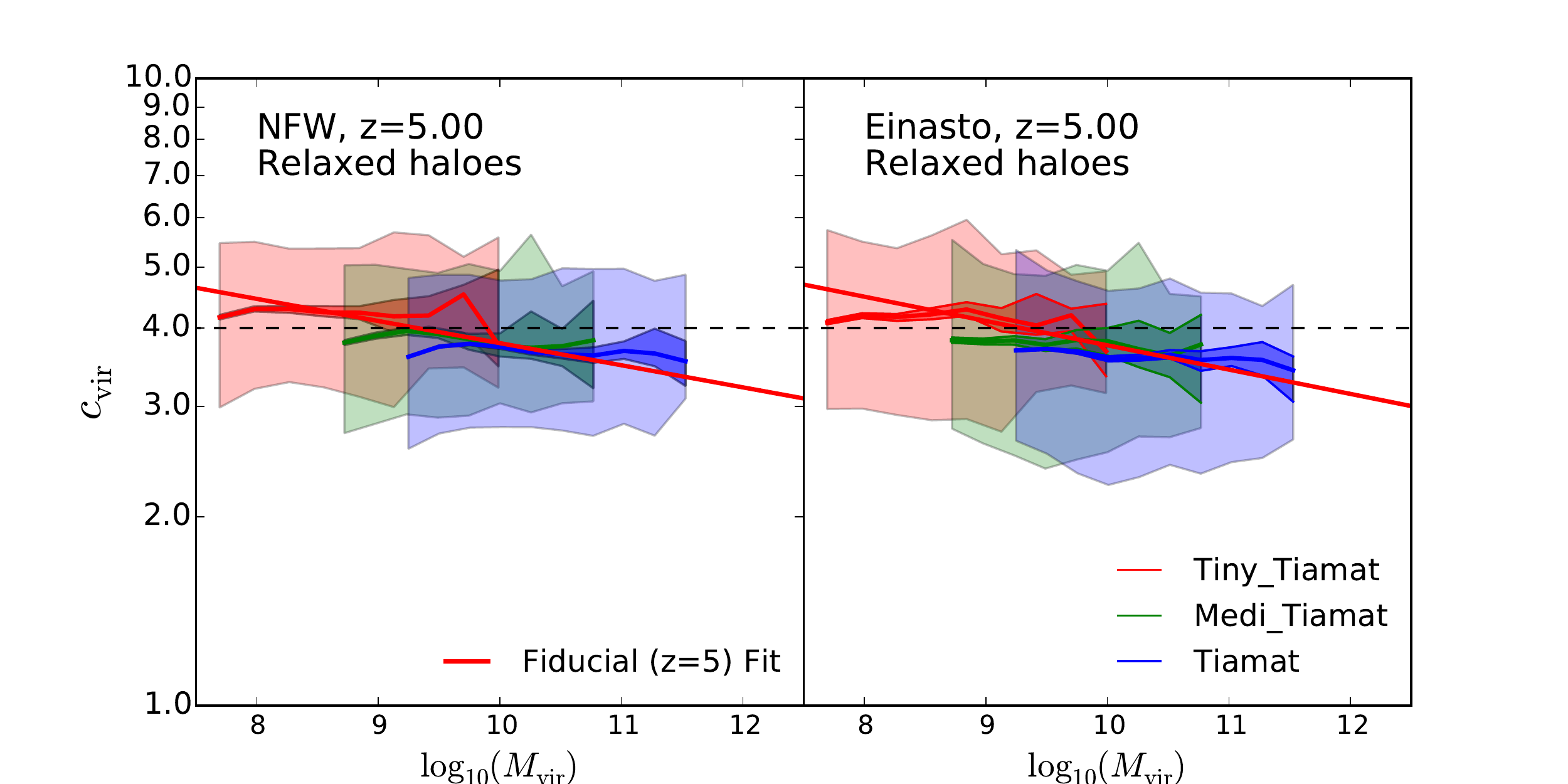}
\caption{Concentration--mass relation for relaxed haloes at $z{=}5$,
  but now with haloes containing ${>}$500 particles. The inner shaded
  region represents the bootstrapped median value and the outer region
  the $68$ per cent scatter. Dashed black line is the $c_{\rm{vir}}=4$ line for reference and the fiducial red line is the best fit to the relaxed population as in Figure \ref{Resolved Concentration z 5}. The inclusion of haloes with particles number ${<}$5000 introduces more haloes with lower concentrations at the low mass end of each simulation. Einasto shape parameters, $\alpha$, for haloes with ${<}$500 particles are also higher as seen in Figure \ref{Full alpha nu}.}
\label{Unresolved concentrations z 5}
\end{center}
\end{minipage}
\end{figure*}

\vspace{7mm}

\vspace{-7mm}
\begin{figure*}
\begin{minipage}{170mm}
\begin{center}
\includegraphics[width=170mm]{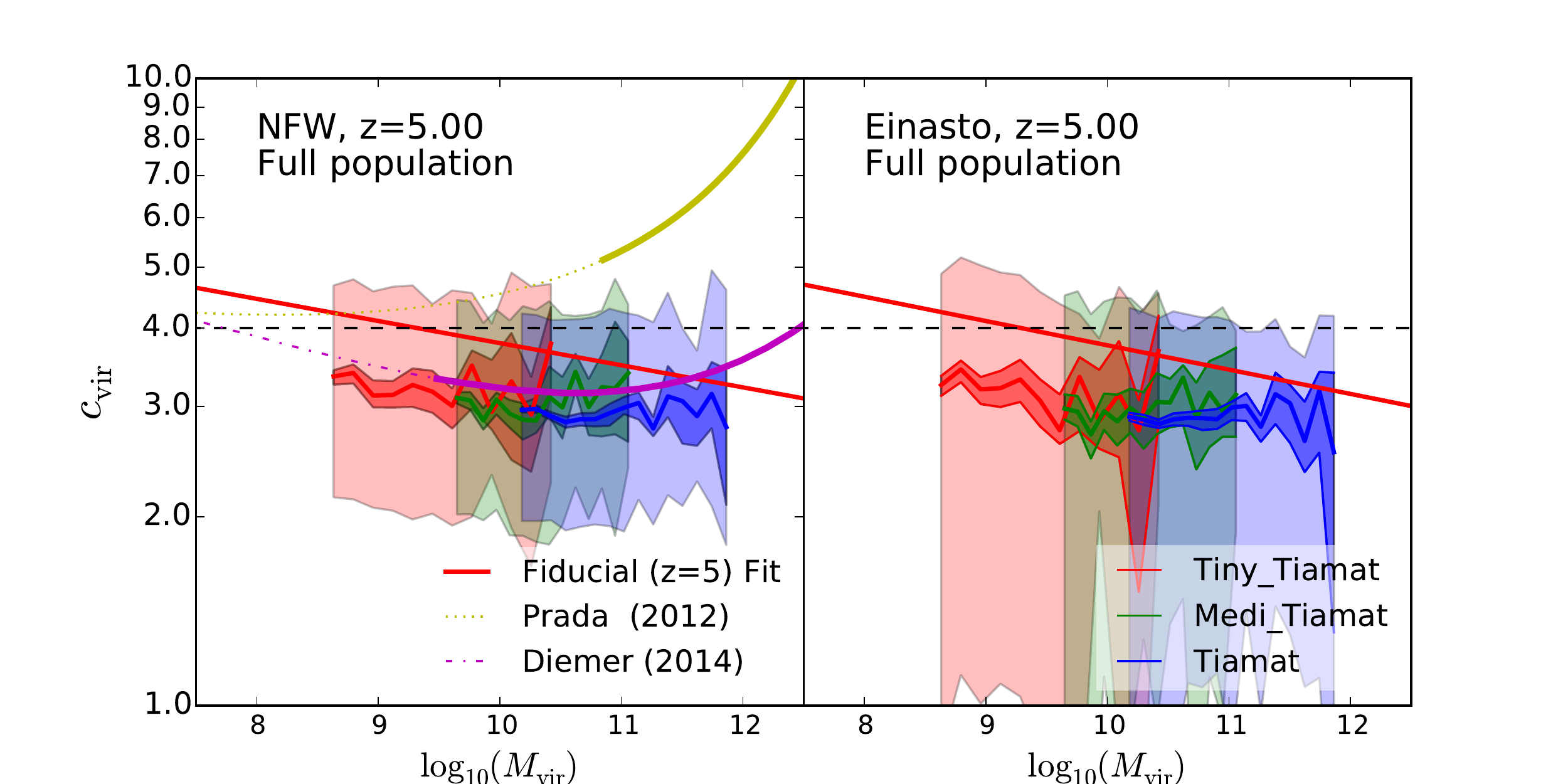}
\caption{The same as Figure \ref{Resolved Concentration z 5} but now the case in which no non-equilibrium cuts are enforced and the full sample of haloes with ${>}$5000 particles is analysed. The inner shaded region represents the bootstrapped median value and the outer region
 the $68$ per cent scatter. The dashed black line at $c_{\rm{vir}}=4$ is added for a point of reference and the fiducial red line is the best fit to the relaxed population as in Figure \ref{Resolved Concentration z 5}. The thick solid line represents previous authors' fits in the mass range their simulations cover, while the thin solid line is the extrapolation to lower masses.  The median magnitude of the concentrations has decreased by $\Delta c_{\rm{vir}}\sim1$ over all masses in our simulations. }
\label{Full Population}
\end{center}
\end{minipage}
\end{figure*}

\subsection{Effects of relaxation and resolution}
\label{subsec:effects of relaxation}

To investigate the effect of our equilibrium selection
criteria we plot in Figure \ref{Concentration vs xoff} the 
variation of concentration with $x_{\rm{off}}$, the most restrictive 
of the three. There is a clear 
trend that haloes with higher $x_{\rm{off}}$ have lower
concentrations, representing a $\Delta c_{\rm{vir}}\sim2$ decrease for 
haloes with the largest offsets (${\sim}30$ per cent of the virial radius). 
The trend is similar at all three of the mass ranges considered. 
This figure illustrates how important it is to understand the 
dynamical state of the haloes included in the sample.

\begin{figure}
\includegraphics[scale=0.4]{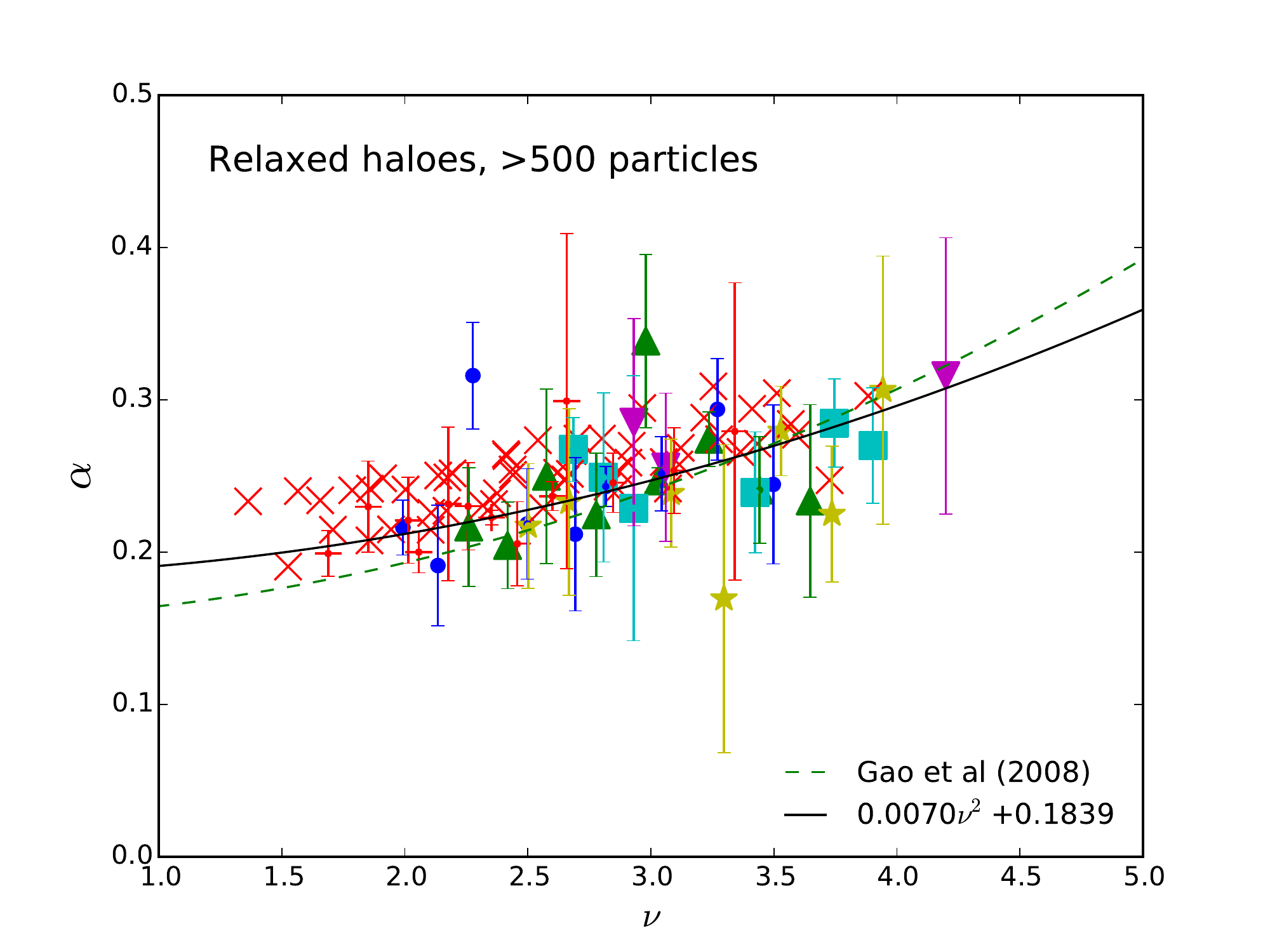}  
\caption{The Einasto $\alpha-\nu$ relation but including haloes above
  $n_{\rm{p}} >500$ particles. At each redshift the median $\alpha$ is
  plotted for bins containing $>20$ haloes, with the errors derived
  from the bootstrapped $90$ per cent confidence interval on the median. Each symbols denotes a different redshift; red pluses denote $z{=}5$, blue filled circles denote $z{=}6$, green triangles denote $z{=}7$, yellow stars denote $z{=}8$, cyan squares denote $z{=}9$ and magenta inverted (point down) triangles denote $z{=}10$. Also, haloes in the range $500 {<} n_{\rm{p}} {<} 5000$ are denoted by the red crosses. These low particle number haloes introduce high $\alpha$ parameters which are not in accord with the best fit quadratics found for more resolved haloes in this and previous works.}
\label{Full alpha nu}
\end{figure}

In Figures \ref{Unresolved concentrations z 5} and 
\ref{Full Population} we again present the
$c({\rm M})$ 
relation but now relax the strict resolution and equilibrium
criteria. Firstly, in Figure \ref{Unresolved concentrations z 5} 
we show the relation that results from lowering the minimum particle 
limit for a halo to 500 particles, while maintaining the relaxation 
criteria. Whereas our $n_{\rm{p}}{>}5000$ relation agreed where 
simulations overlapped, we find a significant discrepancy between 
the simulations for masses corresponding to haloes with $5000>n_{\rm{p}}>500$. In particular, lower 
particle numbers result in lower values of $c_{\rm{vir}}$. For Einasto 
profiles the $\alpha-\nu$ relation also changes slightly, as shown 
in Figure \ref{Full alpha nu}. Many of the $500<n_{\rm{p}}<5000$ haloes 
have $\alpha \sim0.25$ and do not follow the previous quadratic 
$\alpha-\nu$ relation. We thus caution against over-interpreting Einasto fit 
parameters at low particle number. 

In Figure \ref{Full Population} the
$c({\rm M})$ relation is plotted for the {\em entire}
halo sample. At all masses, the 
median concentrations decrease relative to those of our relaxed haloes.
We also note that inclusion of unrelaxed haloes
alters the $\alpha-\nu$ relation slightly. Our best-fit to the full population is $\alpha = 0.0091\nu^2+0.1447$, which is slightly steeper than that found by \citet[][]{2008MNRAS.390L..64D} when they include 
unrelaxed haloes. Table \ref{Full concentration fit parameters} shows 
the best fit parameters for the full population of resolved haloes with $n_{\rm{p}}{>}500$ particles. 

In the left hand panel of Figure \ref{Full Population} we 
also plot the models of \cite{2014arXiv1407.4730D} and
\cite{2012MNRAS.423.3018P}, where we have analytically adjusted the \cite{2012MNRAS.423.3018P} relation from the $c_{200}, M_{200}$ definitions, which they use, to the $c_{\rm{vir}}, M_{\rm{vir}}$ defintions used in this work by assuming an NFW profile. \cite{2014arXiv1407.4730D} use the phase 
space halo finder $\ROCKSTAR$ \citep[][]{2013ApJ...762..109B}, and do 
not enforce any relaxation criteria. They also utilise different 
resolution criteria. The mass range covered in
\cite{2014arXiv1407.4730D} is equivalent to the mass range covered 
by $\Tiamat$ with halo particle numbers of $n_{\rm{p}}{>}1200$. In 
this mass range we see slight evidence for an upturn in the $\Tiamat$
$c({\rm M})$ relation, which is consistent with their 
findings. Note that this upturn is a feature exclusive to our full 
halo sample, suggesting its connection to departures from equilibrium.

\begin{table*}
\begin{minipage}{170mm}
\begin{center}
\begin{tabular}{| l | c  c  c |}
\hline
$z$ & Full Population NFW Best Fit & Full Population Einasto Best Fit & $N_{\rm{sample}}$ \\
\hline
5        & $(3.0{\pm}0.3)(\frac{M}{10^{10}M_{\odot}h^{-1}})^{(-0.026{\pm}0.004)}$ & $(3.0{\pm}0.3)(\frac{M}{10^{10}M_{\odot}h^{-1}})^{(-0.024{\pm}0.005)}$ &18696,3014,3652 \\ 
6        & $(2.8{\pm}0.3)(\frac{M}{10^{10}M_{\odot}h^{-1}})^{(-0.015{\pm}0.004)}$ & $(2.7{\pm}0.4)(\frac{M}{10^{10}M_{\odot}h^{-1}})^{(0.000{\pm}0.006)}$  &10057,1918,2838 \\
7        & $(2.7{\pm}0.3)(\frac{M}{10^{10}M_{\odot}h^{-1}})^{(-0.002{\pm}0.005)}$ & $(2.7{\pm}0.5)(\frac{M}{10^{10}M_{\odot}h^{-1}})^{(0.013{\pm}0.008)}$  &4772,1068,2069\\
8        & $(2.7{\pm}0.4)(\frac{M}{10^{10}M_{\odot}h^{-1}})^{(0.002{\pm}0.007)}$  & $(2.7{\pm}0.7)(\frac{M}{10^{10}M_{\odot}h^{-1}})^{(0.027{\pm}0.011)}$  &1978,565,1347\\
9        & $(2.7{\pm}0.5)(\frac{M}{10^{10}M_{\odot}h^{-1}})^{(0.019{\pm}0.009)}$  & $(2.8{\pm}0.8)(\frac{M}{10^{10}M_{\odot}h^{-1}})^{(0.040{\pm}0.013)}$  &678,268,829 \\
10       & $(2.7{\pm}0.9)(\frac{M}{10^{10}M_{\odot}h^{-1}})^{(0.034{\pm}0.016)}$  & $(2.9{\pm}1.0)(\frac{M}{10^{10}M_{\odot}h^{-1}})^{(0.057{\pm}0.015)}$  &247,142,499 \\
\hline
\end{tabular}
\caption{Best fit values for the full population of NFW and Einasto-derived $c({\rm M})$ relations parameterised with the form $c_{\rm{vir}}$ = $A(\rm{M}/[10^{10}\rm{M}_{\odot}h^{-1}])^B$. $N_{\rm{sample}}$ denotes the number of haloes in the sample for the $\Tiamat$, $\MediTiamat$ and $\TinyTiamat$ simulation respectively.}
\label{Full concentration fit parameters}
\end{center}
\end{minipage}
\end{table*}

Our concentrations at these masses and redshifts are lower than those 
found by \cite{2012MNRAS.423.3018P}. The $\Delta
c_{\rm{vir}}\sim 1$ difference likely originates from a combination of: a)
the use of $M_{200}$ and $c_{200}$ in their definitions of mass and
concentration, b) the use of maximum circular velocity as a measure
of concentration, which is shown in \cite{2013arXiv1303.6158M} to 
reconcile differences in the $c({\rm M})$ relation 
between \cite{2012MNRAS.423.3018P} and \cite{2008MNRAS.390L..64D} (a difference as 
large as $\sim$40 per cent ), c) a different halo finder
(Bound-Density-Maxima), and/or d) a lower particle limit of 500. We 
also note a similar comparison with \cite{2014arXiv1411.4001K}, who 
measure concentrations using circular velocity.
When we relax our halo equilibrium criteria in Figure \ref{Full Population} 
we obtain concentrations at $M = 10^{10.5}M_{\odot}h^{-1}$ which 
are in better agreement with DM14 and \cite{2014arXiv1407.4730D}. 
Our simulations do not show an upturn after unrelaxed haloes are 
removed, as pointed out in \cite{2015arXiv150200391C}.

In the case of both the relaxed population and the full 
population we find the weak trend in the $M_{\rm{vir}}{-}c_{\rm{vir}}$ 
to be steady across this mass range. The redshift dependence of the 
relaxed population of NFW concentrations can be described by 
\begin{equation}
c_{\rm{vir}} {=} 3.2\, \biggl(\frac{{\rm M}}{10^{10}\, {\rm M}_{\odot}h^{-1}}\biggr)^{-0.03}\biggl(\frac{1+z}{10}\biggr)^{-0.29}
\end{equation}
and Einasto concentrations by
\begin{equation}
c_{\rm{vir}} {=} 3.1\, \biggl(\frac{{\rm M}}{10^{10}\, {\rm M}_{\odot}h^{-1}}\biggr)^{-0.03}\biggl(\frac{1+z}{10}\biggr)^{-0.29}.
\end{equation}
The full population of NFW concentrations is described by
\begin{equation}
c_{\rm{vir}} = 2.6\, \biggl(\frac{{\rm M}}{10^{10}\, {\rm M}_{\odot}h^{-1}}\biggr)^{-0.03}\biggl(\frac{1+z}{10}\biggr)^{-0.21}
\end{equation}
and Einasto concentrations by
\begin{equation}
c_{\rm{vir}} = 2.6\, \biggl(\frac{{\rm M}}{10^{10}\, {\rm M}_{\odot}h^{-1}}\biggr)^{-0.03}\biggl(\frac{1+z}{10}\biggr)^{-0.11}.
\end{equation}
Here we assume the slope is constant across the mass range and only
fit to the normalisation of the $c_{\rm{vir}}{-}M_{\rm{vir}}$
relation. We also note that the trend for both NFW and Einasto 
fits are consistent with being mass-independent at $z>8$, as seen in 
Table \ref{Relaxed concentration fit values}.

\vspace{-7mm}
\section{Spin Parameter}
\begin{figure}
\includegraphics[scale=0.4]{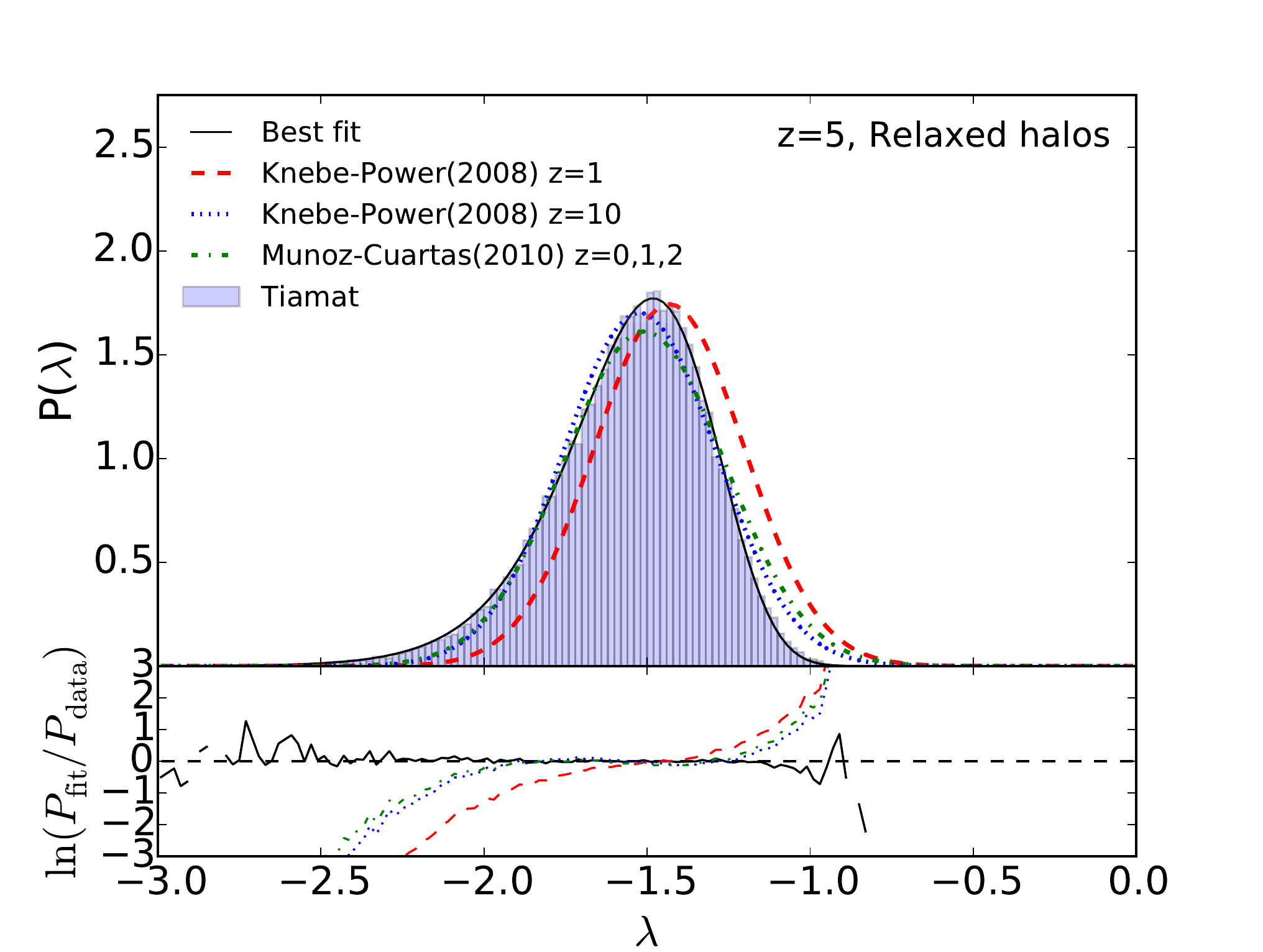}
\caption{The distribution of Bullock spin parameters for relaxed haloes at redshift 5 in $\Tiamat$. The best fit to Eq \ref{Bett spin distribution} centered on $\lambda_0 = 0.033$ and $\alpha = 2.25$ is shown as the solid line. Fits from several previous studies are also shown. The lower panel shows the fractional distribution relative to the simulation.}
\vspace{-2mm}
\label{Relaxed spin distribution z 5}
\end{figure}

In Figure \ref{Relaxed spin distribution z 5} we show the distribution
of spin parameters for relaxed haloes in the $\Tiamat$ simulation;
results are shown for $z=5$. The mass range covered is now increased so that 
$n_{\rm{p}} > 600$ \citep[as in][]{2008ApJ...678..621K}. Rather than a log-normal, the solid black lines
show the best-fitting function of \cite{2007MNRAS.376..215B}:

\begin{equation}
P(\lambda) = \frac{3\ln(10)}{\Gamma(\alpha)}\alpha^{\alpha-1} \biggl(\frac{\lambda}{\lambda_0}\biggr)^3\exp\biggl[-\alpha \biggl(\frac{\lambda}{\lambda_0}\biggr)^{3/\alpha}\biggr].
\label{Bett spin distribution}
\end{equation}

This function has been shown to better describe the low spin 
tail. Also shown are log-normal fits by \cite{2008ApJ...678..621K} and 
\cite{2010crf..work...16M}. Although evaluated at different redshifts 
from $\Tiamat$, these authors report minimal redshift-evolution in the halo spins. 
Our results support this, with the
distribution being well fit by Eq \ref{Bett spin distribution} with 
a small dependence on redshift. At $z{=}5$ we have  
$(\lambda_0, \alpha){=}(0.033\pm0.0002, 2.25 \pm 0.04)$ 
changing to $(\lambda_0, \alpha){=}(0.029 {\pm} 0.0004, 2.36{\pm}0.1)$ 
at $z{=}10$. Best fit parameters and errors are again derived using 
the MCMC method described in Section 3. Parameters for all redshifts 
are shown in Table \ref{Relaxed spin distribution parameters} along 
with the numbers of haloes in each sample.

\begin{figure}
\includegraphics[scale=0.4]{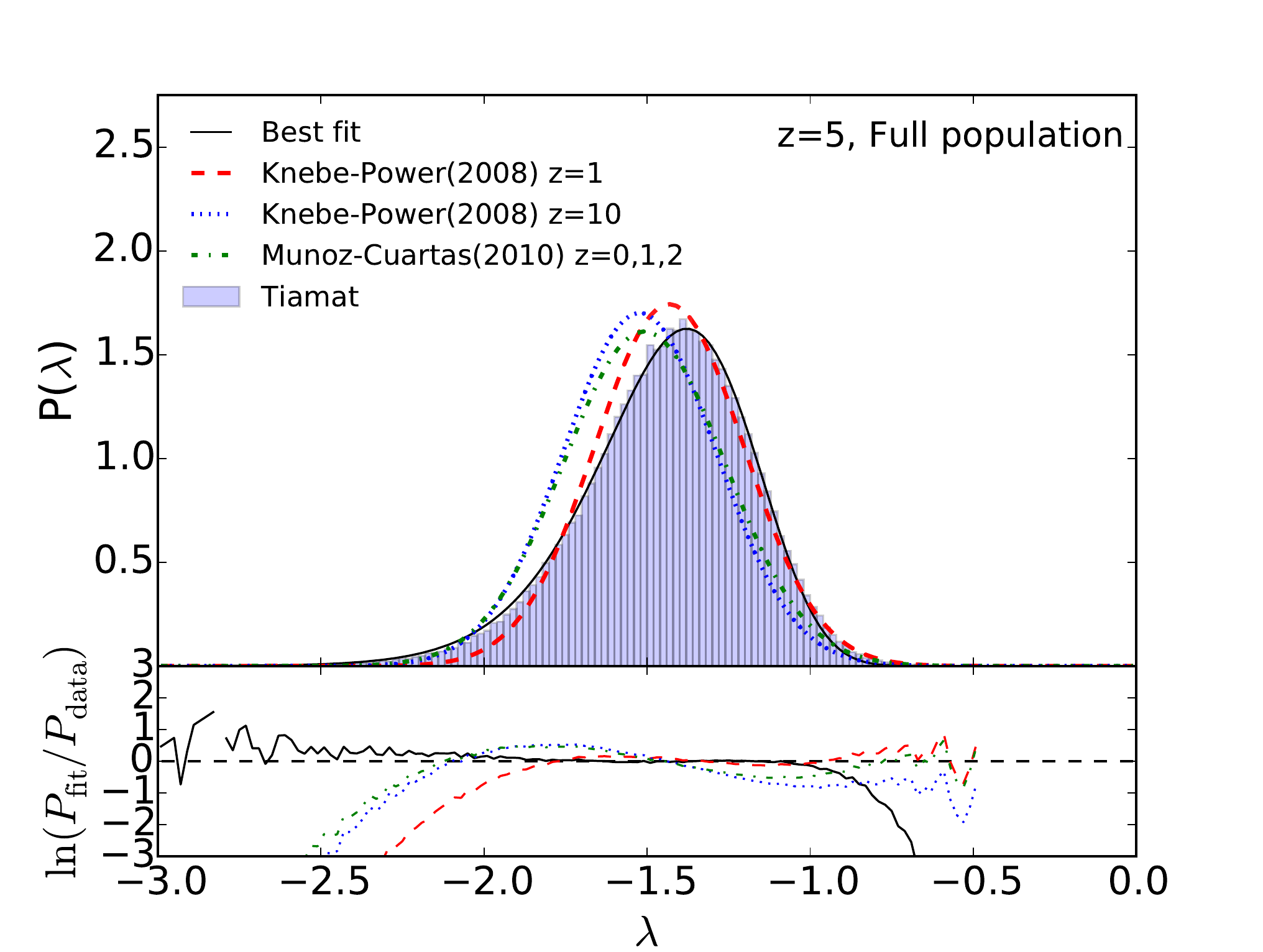}
\caption{The distribution of Bullock spin parameters for the full sample of haloes in $\Tiamat$ at redshift 5, without cutting unrelaxed haloes. The best fit to Eq \ref{Bett spin distribution} is centered on $\lambda_0 = 0.042$ and $\alpha = 2.70$. Fits from several previous studies are also shown. The lower panel shows the fractional distribution relative to the simulation.}
\vspace{-2mm}
\label{Full spin distribution z 5}
\end{figure}

Both \cite{2010crf..work...16M} and \cite{2008ApJ...678..621K}
fit a log-normal to their spin distribution, with best-fit parameters
given by $\sigma_0= 0.57$ (variance) and $\lambda_0 = 0.031$ (mean), and 
$\sigma_0 = 0.53$ and $\lambda_0 =0.035$, respectively. 
Both have slightly higher spins overall, and a more extended tail at the
upper end of the distribution. As found by \cite{2007MNRAS.376..215B}, 
eq. \ref{Bett spin distribution} provides a better fit to our low spin 
distribution than does the log-normal. We find that 
unrelaxed haloes have a noticeable impact on our spin
distribution, which is shown in Figure \ref{Full spin distribution z 5}. 
Without removing these haloes we find the best-fit parameters to be 
$(\lambda_0, \alpha) =( 0.042 {\pm} 0.0002, 2.70{\pm} 0.04)$. 

\begin{figure}
\includegraphics[scale=0.4]{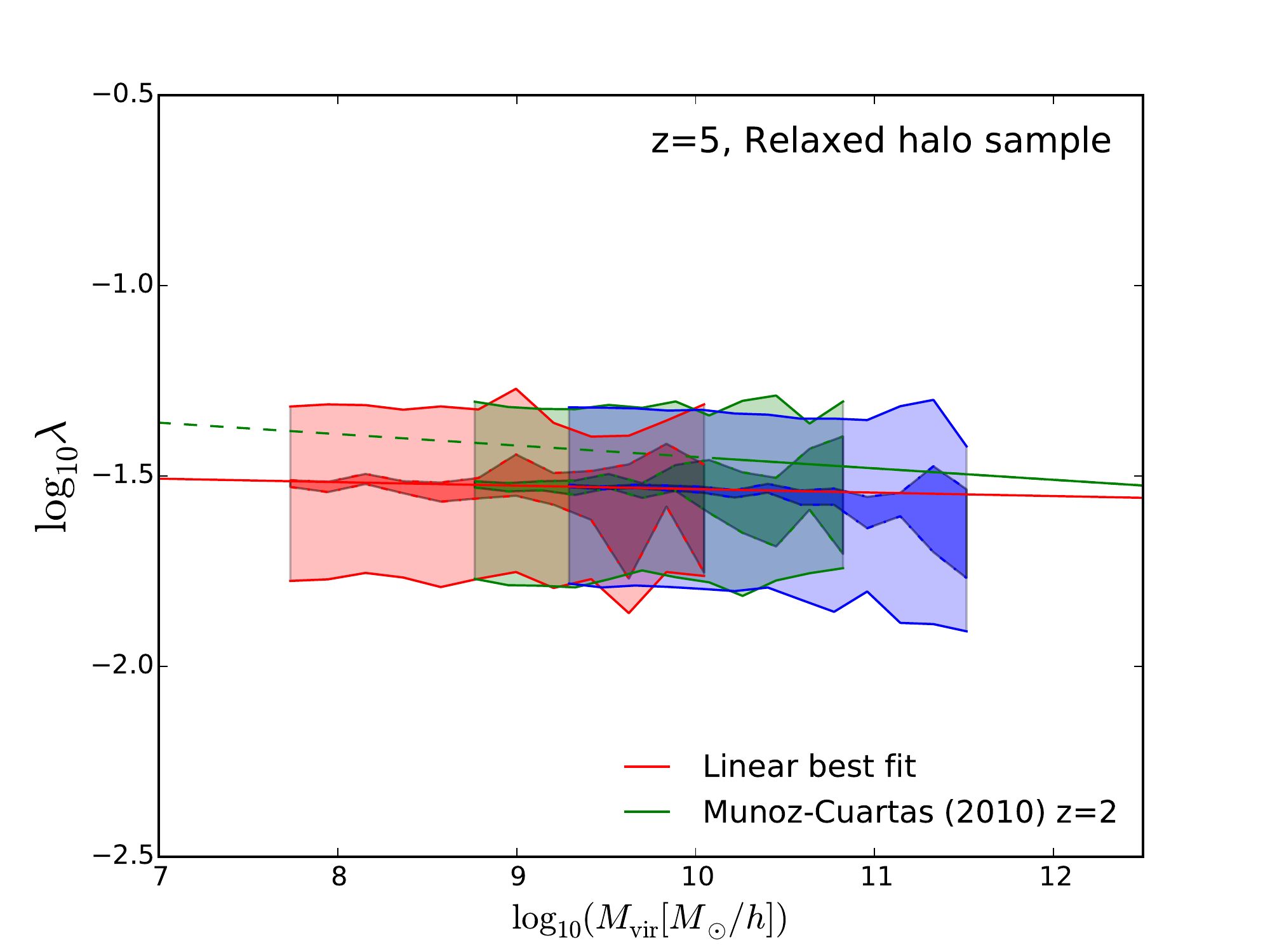}
\caption{The spin-virial mass relation for relaxed haloes at
  $z{=}5$. Inner shaded region denotes the bootstrapped 90 per cent
  confidence interval on the median. The outer shaded region shows the
  $68$ per cent scatter. The colours red, green and blue represent the $\TinyTiamat$, $\MediTiamat$ and $\Tiamat$ simulations. For comparison results for relaxed haloes from \protect\cite{2011MNRAS.411..584M} are also shown.}
\label{Relaxed spin mass z 5}
\end{figure}

Figure \ref{Relaxed spin mass z 5} shows the relation between 
spin and virial mass at $z=5$, with a power-law best-fit of the form

\begin{equation}
\lambda = A\biggl(\frac{M_{\rm{vir}}}{10^{10}\, {\rm M}_{\odot}h^{-1}}\biggr)^B.
\end{equation}
A slight negative slope of $B{=}-0.01 \pm 0.006$ at redshift 
$z{=}5$ decreases to $-0.023 \pm 0.016$ 
by $z{=}10$. Fits from previous work that study the Bullock spin 
parameter at high redshift (Knebe-Power at $z{=}10$ and Munoz-Cuartas 
at $z{=}2$) are also shown. We find the scatter in the spin to be 
roughly constant in each mass bin. Parameter fits for all our
redshifts are provided in Table \ref{Spin mass parameters} for relaxed 
haloes as well as the full population.

\begin{table}
\begin{center}
\begin{tabular}{| l | c  c  c |}
\hline
$z$ & $\lambda_0$ & $\alpha$ & $N_{\rm{sample}}$ \\
\hline
5        & 0.033  $\pm$ 0.0002 & 2.25 $\pm$ 0.04 & 61857  \\ 
6        & 0.032  $\pm$ 0.0002 & 2.30 $\pm$ 0.05 & 41035  \\
7        & 0.031  $\pm$ 0.0002 & 2.34 $\pm$ 0.06 & 25384  \\
8        & 0.030  $\pm$ 0.0003 & 2.30 $\pm$ 0.07 & 13996  \\
9        & 0.030  $\pm$ 0.0003 & 2.30 $\pm$ 0.08 & 6607   \\
10       & 0.029  $\pm$ 0.0004 & 2.36 $\pm$ 0.10 & 3393   \\
\hline
\end{tabular}
\caption{Best fit values for distribution of relaxed population spin parameters according to Equation \ref{Bett spin distribution}. $N_{\rm{sample}}$ is the number of haloes in the sample from $\Tiamat$.}
\label{Relaxed spin distribution parameters}
\end{center}
\end{table}

\begin{figure}
\includegraphics[scale=0.4]{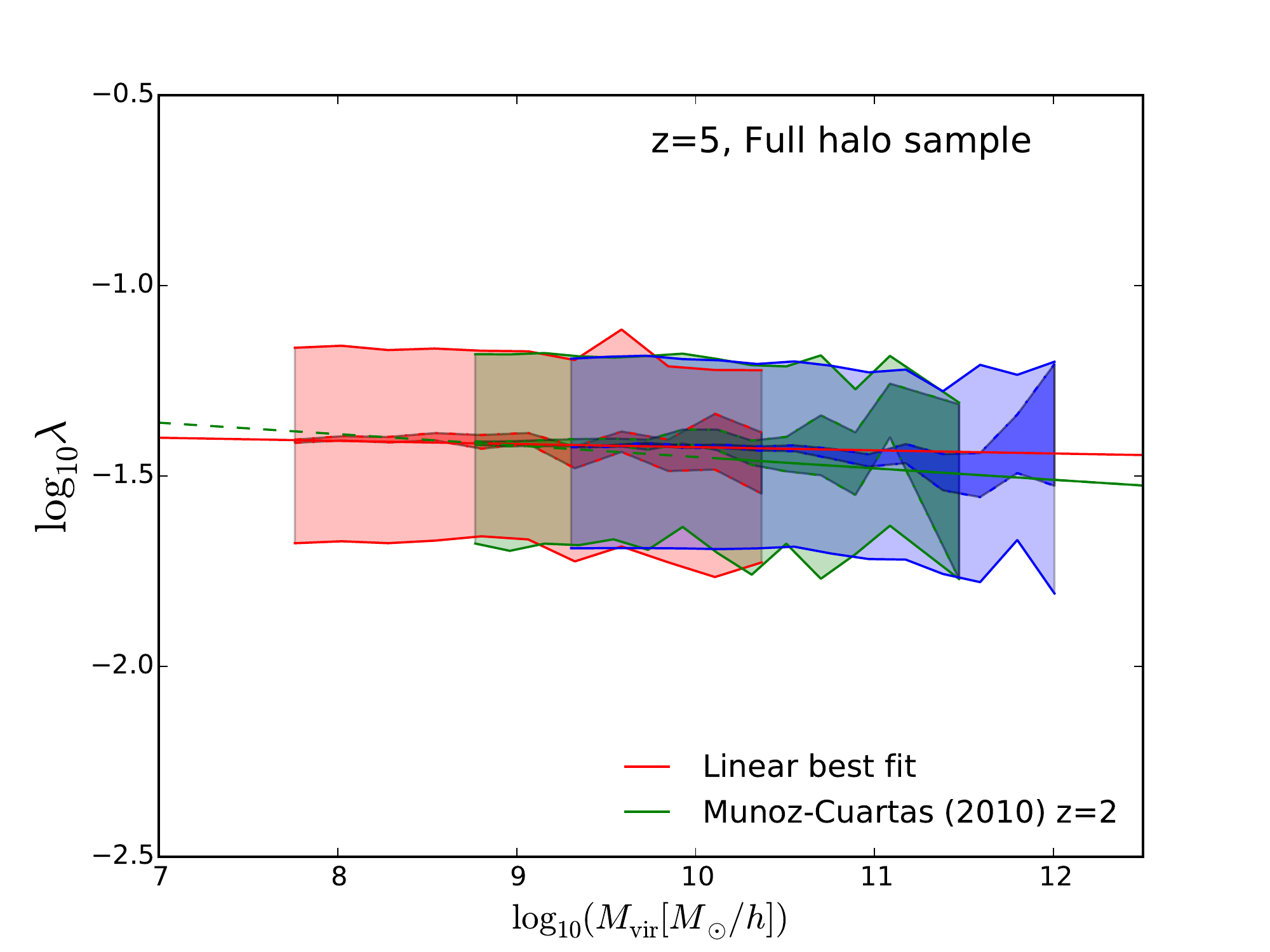}
\caption{The spin-virial mass relation for the full population of
  haloes at $z{=}5$. Inner shaded region denotes the bootstrapped 90 per cent
  confidence interval on the median. The outer shaded region shows the
  $68$ per cent scatter. The colours red, green and blue represent the $\TinyTiamat$, $\MediTiamat$ and $\Tiamat$ simulations. For comparison fiducial results for relaxed haloes from \protect\cite{2011MNRAS.411..584M} are also shown.}
\label{Full spin mass z 5}
\end{figure}

The existence of a small negative trend of spin parameter with 
mass at these redshifts is in qualitative agreement with
\cite{2008ApJ...678..621K}, 
who find no trend at $z{=}0-1$ but an emerging trend at $z{=}10$. As 
noted, the virialised halo cut has an affect on our results. However, 
we find a small negative trend in both the full and relaxed
population. For example, the $\lambda - M_{\rm{vir}}$ relation for 
the full sample at redshift 5 is shown in Figure \ref{Full spin mass z 5}. 
Without making cuts we find a relation with slope $A{=}$-0.009 at
$z{=}5$, changing to $B{=}$-0.029 at redshift 10. Our spin mass
relation with sample cuts is in agreement with the result of 
$B{=}-0.06\pm0.17$ from \cite{2008ApJ...678..621K}.

\begin{table*}
\begin{minipage}{170mm}
\begin{center}
\begin{tabular}{|l|cccc|}
\hline
$z$ & Relaxed Haloes Best Fit & $N_{\rm{relaxed}}$ & Full Population Best Fit & $N_{\rm{total}}$\\
\hline
5 & $(0.029{\pm}0.004)(\frac{M}{10^{10}M_{\odot}/h})^{(-0.009{\pm}0.006)}$ &61854,8510,8539& $(0.038{\pm}0.002)(\frac{M}{10^{10}M_{\odot}/h})^{(-0.009{\pm}0.003)}$ & 234992,32174,32208\\ 
6 & $(0.028{\pm}0.004)(\frac{M}{10^{10}M_{\odot}/h})^{(-0.013{\pm}0.007)}$ &41034,6156,6843& $(0.036{\pm}0.003)(\frac{M}{10^{10}M_{\odot}/h})^{(-0.014{\pm}0.004)}$ & 163111,24791,27889\\
7 & $(0.028{\pm}0.005)(\frac{M}{10^{10}M_{\odot}/h})^{(-0.016{\pm}0.008)}$ &25384,4258,5381& $(0.035{\pm}0.003)(\frac{M}{10^{10}M_{\odot}/h})^{(-0.020{\pm}0.004)}$ & 104335,17832,22982\\
8 & $(0.027{\pm}0.006)(\frac{M}{10^{10}M_{\odot}/h})^{(-0.021{\pm}0.010)}$ &13996,2673,4094& $(0.034{\pm}0.004)(\frac{M}{10^{10}M_{\odot}/h})^{(-0.021{\pm}0.005)}$ & 61194,12049,18065 \\
9 & $(0.025{\pm}0.007)(\frac{M}{10^{10}M_{\odot}/h})^{(-0.026{\pm}0.013)}$ &6607,1519,2832 & $(0.033{\pm}0.004)(\frac{M}{10^{10}M_{\odot}/h})^{(-0.027{\pm}0.006)}$ & 31159,7194,13227  \\
10& $(0.025{\pm}0.009)(\frac{M}{10^{10}M_{\odot}/h})^{(-0.023{\pm}0.016)}$ &3393,836,1980  & $(0.032{\pm}0.005)(\frac{M}{10^{10}M_{\odot}/h})^{(-0.029{\pm}0.008)}$ & 16132,4303,9621   \\
\hline
\end{tabular}
\caption{Best fit values spin-mass relation for relaxed, and the full population of haloes, $\lambda = A(\rm{M}/[10^{10}M_{\odot}h^{-1}])^B$. $N_{\rm{relaxed(total)}}$ denotes the number of haloes in the relaxed (full population) sample for the $\Tiamat$, $\MediTiamat$ and  $\TinyTiamat$ simulation respectively. }
\label{Spin mass parameters}
\end{center}
\end{minipage}
\end{table*}

\vspace{-7mm}
\section{Discussion and Conclusion}
We used N-Body simulations to study concentrations and spins of 
DM haloes at $z{=}5{-}10$ and across the mass range $10^8M_{\odot}h^{-1} < M <10^{11}M_{\odot}h^{-1}$; 
the regime relevant for studies of structure formation during the epoch 
of reionization. The dependence of these parameters on 
equilibrium state was investigated by splitting our 
halo sample into two populations which include i) only relaxed haloes 
and ii) the full population. We find qualitatively similar 
results to previous studies. However, we find quantitative 
differences between our derived $c({\rm M})$ relations and spin-mass 
relations and those of previous studies, which we attribute to
each author's use of different halo finders, to our higher simulation resolution, 
and to the different relaxation criteria used for our sample. We find the model 
proposed by \citet{2014MNRAS.441..378L} reproduces both the slope and 
redshift evolution of our $c({\rm M})$ relation.

Our key results are as follows:
\begin{itemize}

\item Our best-fit concentration--mass relations at $z{=}5$ 
have a slightly negative slope that 
becomes more shallow towards $z{=}9$. Limiting our analysis to
equilibrium haloes has a strong impact on the derived $c({\rm M})$
relation due to unrelaxed haloes having lower concentrations at all masses 
and redshifts. Haloes with larger center-of-mass offset 
($x_{\rm{off}}$) typically have lower concentrations. 

\item The slope of the $c({\rm M})$ relation becomes 
shallower at higher redshift, although concentrations decrease 
at all masses. However, at high redshifts the number of haloes 
passing the equilibrium criteria is low: only $\sim30$ per cent of haloes 
in the 67.8 Mpc$/h$ box pass our resolution and relaxation cuts at 
$z=5$. Such a high proportion of unrelaxed haloes at the mass scales 
studied here is a distinct property of the high redshift universe, as 
discussed in Paper I. We find concentrations of relaxed haloes at 
$z{>}5$ to be well described by the relation 
$c_{\rm{vir}} {=}3.2\, ({\rm M}/[10^{10}M_{\odot}h^{-1}])^{-0.03}[(1+z)/10]^{-0.29}$ 
for NFW fits and 
$c_{\rm{vir}} {=}3.1\, ({\rm M}/[10^{10}M_{\odot}h^{-1}])^{-0.03}[(1+z)/10]^{-0.29}$ 
for Einasto concentrations. The intrinsic scatter around the 
$c({\rm M})$ relations is $\Delta c_{\rm{vir}}\sim1$ 
(or 20 per cent). We find the shape parameter of the Einasto profiles to 
depend on the peak height mass parameter approximate as $\alpha = 0.007\nu^2+0.184$.

\item Without imposing equilibrium cuts on our sample, the concentrations found in 
$\Tiamat$ have similar values to those reported by 
\cite{2014arXiv1402.7073D}, \cite{2014arXiv1407.4730D} and 
\cite{2015arXiv150506436H}. Concentrations of haloes in $\Tiamat$ are
a factor of $\Delta c_{\rm{vir}} \sim 0.5-1$ lower than reported by 
\cite{2012MNRAS.423.3018P} and \cite{2014arXiv1411.4001K}. The shallow 
negative trend in the $c({\rm M})$ relation that 
flattens from $z=5$ to $z=10$, and the overall decrease in the
magnitude of our Einasto concentrations agree well with \cite{2015arXiv150506436H}.

\item The distribution of Bullock spin parameters for relaxed haloes 
at $z {\geq} 5$ is found to be well fit by eq.~\ref{Bett spin distribution}
with $(\lambda_0, \alpha){\sim}(0.033\pm0.0002, 2.25\pm0.04)$, with
little evolution with redshift. Including 
unrelaxed haloes results in a spin distribution with a higher mean
of $\lambda_0{=}0.042$. 

\item As in previous studies, we find a spin-virial mass relation with 
a slight negative correlation at high redshift. The trend found here
has a slope $d\log\lambda/d\log {\rm M} \sim -0.02$ at 
$z{=}10$. The exclusion of unrelaxed haloes also has the effect of 
increasing the peak of the spin distribution while the slope of the 
$\lambda-M_{\rm{vir}}$ relation remains slightly negative. Our
best-fit power-law relation relaxed haloes at $z{=}5$ is 
$\lambda =(0.029\pm0.004)({\rm M}/[10^{10}M_{\odot}h^{-1}])^{-0.009\pm0.006}$ 
and $\lambda =(0.038\pm0.002)({\rm M}/[10^{10}M_{\odot}h^{-1}])^{-0.009\pm0.003}$ 
for the full halo population.

\end{itemize}

The growth of dark matter haloes drives high-$z$ galaxy
formation \citep[][]{2013ApJ...768L..37T}, while the concentration 
and spin of haloes are key ingredients for semi-analytic models of 
galaxy formation \citep[e.g.][]{2006MNRAS.365...11C}. This study of 
these properties for haloes corresponding to the galaxies responsible 
for reionization will provide a valuable resource for understanding 
the framework of early galaxy formation.
\\
\\

{\bf Acknowledgements}
This research was supported by the Victorian Life Sciences Computation
Initiative (VLSCI), grant ref. UOM0005, on its Peak Computing Facility
hosted at the University of Melbourne, an initiative of the Victorian
Government, Australia. Part of this work was performed on the gSTAR
national facility at Swinburne University of Technology. gSTAR is
funded by Swinburne and the Australian GovernmentÕs Education
Investment Fund. This research program is funded by the Australian
Research Council through the ARC Laureate Fellowship FL110100072
awarded to JSBW. ADL is financed by a COFUND Junior Research
Fellowship. We thank Volker Springel for making the GADGET2 and 
SUBFIND codes available. We also thank N.Gnedin for useful comments 
on our manuscript.

\vspace{-5mm}
\newcommand{\noopsort}[1]{}
\bibliographystyle{mn2e}
\bibliography{bibliography}
\label{lastpage}
\end{document}